\documentclass[letterpaper,english,preprint,aps,nofootinbib]{revtex4-1}
\pdfoutput=1
\usepackage[T1]{fontenc}
\usepackage{xcolor}
\usepackage{amsmath}
\usepackage{amssymb}
\usepackage{graphicx}
\usepackage{babel}

\makeatletter

\pdfpageheight\paperheight
\pdfpagewidth\paperwidth

\usepackage{babel}

\@ifundefined{showcaptionsetup}{}{%
\PassOptionsToPackage{caption=false}{subfig}}
\usepackage{subfig}
\makeatother

\begin{document}

\title{Generation of vortices and stabilization of vortex lattices in holographic
superfluids}

\author{Xin Li}
\email{lixin615@mails.ucas.edu.cn}
\affiliation{School of Physical Sciences, University of Chinese Academy of Sciences,
Beijing 100049, China}

\author{Yu Tian}
\email{ytian@ucas.ac.cn}
\affiliation{School of Physical Sciences, University of Chinese Academy of Sciences,
Beijing 100049, China}
\affiliation{Institute of Theoretical Physics, Chinese Academy of Sciences, Beijing
100190, China\\
Center for Gravitation and Cosmology, College of Physical Science
and Technology, Yangzhou University, Yangzhou 225009, China}

\author{Hongbao Zhang}
\email{hzhang@vub.ac.be}
\affiliation{Department of Physics, Beijing Normal University, Beijing 100875,
China}
\affiliation{Theoretische Natuurkunde, Vrije Universiteit Brussel, and The International
Solvay Institutes, Pleinlaan 2, B-1050 Brussels, Belgium}

\date{\today}
\begin{abstract}
Within the simplest holographic superfluid model and without any ingredient
put by hand, it is shown that vortices can be generated when the angular
velocity of rotating superfluids exceeds certain critical values,
which can be precisely determined by linear perturbation analyses
(quasi-normal modes of the bulk AdS black brane). These vortices appear
at the edge of the superfluid system first, and then automatically
move into the bulk of the system, where they are eventually stabilized
into certain vortex lattices. For the case of 18 vortices generated,
we find (at least) five different patterns of the final lattices formed
due to different initial perturbations, which can be compared to the known result for such lattices in weakly coupled Bose-Einstein condensates from free energy analyses. 
\end{abstract}

\maketitle
\global\long\def\cL{\mathcal{L}}%
\global\long\def\cd{\cdots}%
\global\long\def\d{\delta}%
\global\long\def\i{\int}%
\global\long\def\ka{\kappa}%
\global\long\def\l{\lambda}%
\global\long\def\m{\mu}%
\global\long\def\na{\nabla}%
\global\long\def\o{\omega}%
\global\long\def\Om{\Omega}%
\global\long\def\p{\partial}%
\global\long\def\r{\rho}%
\global\long\def\s{\psi}%
\global\long\def\S{\Psi}%
\global\long\def\t{\theta}%

\section{Introduction}

Vortices play a central role in the non-trivial dyamics of superfluids. The study of vortex generation and vortex lattice formation in rapidly rotating superfluids, both experimentally and theoretically, has a long history, from that in the old liquid helium-4 or helium-3 to that in the modern Bose-Einstein condensates (BEC) or superfluid ultracold fermions. Quantized vortices and vortex lattices are firstly discovered in helium experiments where the superfluid is driven by a rotating bucket\cite{Helium_vortices,Helium_lattices} (see \cite{R-books} for a review). The interest for vortices in BEC comes from the fact that the theories are more tractable and size of vortex core in BEC is much larger than that in liquid helium\cite{BEC_Crtical_Omega}. As a result, a large quantity of experiments on vortices and vortex lattices are performed in BEC (see \cite{Fetter-RMP} for a review), among which the critical angular velocity is determined in \cite{BEC_Crtical_Omega} and formation of lattices with large numbers of vortices are performed in \cite{BEC_Lattices_Laser,BEC_Lattices_Cooling}. Most of the related theoretical works are based on the Gross-Pitaevskii (GP) equation\cite{GP-G-1961,GP-P-1961,GP-1963} as well as its generalizations, which is a mean field description of weakly coupled BEC at zero temperature, for example, formations of vortex lattices are studied in \cite{GP-2002-1,GP-2002-2,GP-2016}.  

The GP equation based method is relatively simple, and can reveal some features of vortex generation and vortex lattice formation observed in experiments. In \cite{GP-2002-2}, for example, after entering the BEC, vortices move towards the central area of the rotating condensate and finally form a regular lattice, which can reveal the basic features of the lattice formation process. However, it is not difficult to find several disadvantages in these GP equation based methods. First of all, there is no dissipation in the original GP equation, so in order to yield the dissipation effect at finite temperature, a classical thermal cloud has to be introduced\cite{GP-2002-2}. In the numerical simulations by this method, vortices seem to nucleate in the thermal cloud and then enter the BEC to form the lattice. But an explanation how the classical vortices transform to the quantum objects is absent. Second, the GP equation can describe the weakly coupled BEC well, but is not expected
to be a proper desciption of liquid helium systems, which have strong interparticle interactions, as well as general ultracold fermion systems.\footnote{The GP equation can be a good desciption of superfluid ultracold fermion systems in their BEC limit\cite{PS}.}

It is thus valuable to look for alternative methods to cope with the
problem of vortex generation and vortex lattice formation. Applied
AdS/CFT duality, or holography, just provides one of such alternative
methods, which has been proved as a very useful framework to study
superfluid physics at finite temperature (see e.g. \cite{Dissipation and Energy Flow,GKLTZ}).
Holography equates certain quantum systems without gravity to gravitational
systems in a curved spacetime with one additional spatial dimension
\cite{Maldcena,GKPW-GKP,GKPW-W}. It describes strongly coupled systems
by design and naturally incorporates dissipations by including a black
hole on the gravity side. Most interestingly, even the simplest holographic
superfluid model\cite{HHH Model} shows some features of superfluid
ultracold fermions \cite{BEC-BCS,WuChaoLun}. All the above facts
make holography a very promising tool to describe superfluids other
than the zero temperature weakly coupled BEC. In fact, a number of holographic studies of the vortex are performed, where single vortex solutions are considered in \cite{Vortices_PRL,Vortices_JHEP_2010,Vortices_PRD,Vortices_JHEP_2014}, and vortex lattice solutions obtained from static equations can be found in \cite{Lattice_PRD,Lattice_Backreaction_1,Lattice_Backreaction_2}.

In this paper, we use the simplest holographic superfluid model to
study the vortex generation and vortex lattice formation in rapidly
rotating superfluids. Compared with a requisite thermal cloud introduced in \cite{GP-2002-2,GP-2016} or other dissipation terms in \cite{GP-2002-1}, no additional ingredient is required in our model. Actually, we first investigate the critical angular velocity from the linear instability indicated by critical quasi-normal modes (QNM) of the bulk AdS black brane, which is a good comparison with experimental work \cite{BEC_Crtical_Omega} and theoretical work \cite{BEC_Crtical_Omega_Theoretics_PRL}. Next, in the nonlinear time evolution, it is shown that the vortices appear at the edge of the superfluid system at first, and then automatically move into the bulk of the system, where they are eventually stabilized into certain vortex lattices. In particular, we carefully examine the final configurations when there are 18 vortices generated in total. In this case, we find (at least) five different patterns of the final lattices formed due to different initial perturbations, which can be compared to the known result for such lattices in weakly coupled Bose-Einstein condensates from free energy analyses. Moreover, the influences of radius, temperature and angular velocity on both the linear analysis and nonlinear evolutions are discussed. Experimentally, the setup in \cite{BEC_Lattices_Cooling} is most relevant to our theoretical study, where superfluids already have angular velocities at the initial stage when the condensate forms.

This paper is organized as follows. In the next section, we briefly
introduce the holographic superfluid model that will be used. In Sec.~\ref{sec:QNM},
we construct our initial configurations and consider the linear instability
from the QNM on this initial background. Then, in Sec.~\ref{sec:Nonlinear},
the dynamical process of vortex generation and vortex lattice formation
is studied numerically by the nonlinear time evolution, and the final
vortex lattices with 18 vortices are investigated. Finally, the conclusion
with some discussion provided in the last section.

\section{Holographic Setup}

\subsection{Action of the model}

The simplest holographic model to describe superfluids is given in
\cite{HHH Model}, which consists of a complex scalar field $\Psi$
coupled to a $U(1)$ gauge field $A_{M}$ in the $(3+1)D$ gravity with
a cosmological constant related to the AdS radius as $\Lambda=-3/L^{2}$. The action is: 
\begin{equation}
S=\i_{M}\sqrt{-g}d^{4}x\left[\frac{1}{2\kappa^{2}}\left(R+\frac{6}{L^{2}}\right)-\frac{1}{e^{2}}\left(\frac{1}{4}F^{2}+\left|D\S\right|^{2}+m^{2}\left|\S\right|^{2}\right)\right],
\end{equation}
where $D_{M}\S=\left(\na_{M}-iA_{M}\right)\S$, and $\ka^{2}=8\pi G$
with $G$ Newton's gravitational constant.

A simple way to solve the model is considering the probe limit, where
backreaction of the matter fields is neglected. To do so, we take
the limit $e\to\infty$. As a result, the matter fields live in a
bulk spacetime that is fixed to be the Schwarzschild-AdS balck brane
\begin{equation}
ds^{2}=\frac{L^{2}}{z^{2}}\left(-f\left(z\right)dt^{2}+\frac{1}{f\left(z\right)}dz^{2}+dx^{2}+dy^{2}\right),\label{eq:Sch-AdS}
\end{equation}
for the finite temperature system, where $f\left(z\right)=1-(\frac{z}{z_{h}})^{3}$
is the blackening factor. The corresponding
Hawking temperature is 
\begin{equation}
T=\frac{3}{4\pi z_{h}},
\end{equation}
which is identified with the temperature for
the boundary system. The equations of motion are 
\begin{align}
D^{M}D_{M}\S-m^{2}\S & =0, \label{eq:K-G}\\
\na_{M}F^{MN} & =j^{N}, \label{eq:Maxwell}
\end{align}
with $j^{N}=i\left(\S^{*}D^{N}\S-\S D^{N*}\S^{*}\right)$ .

Without losing generality, we choose $m^{2}L^{2}=-2$, so the asymptotic
behaviors of the matter fields are 
\begin{align}
\S & =\frac{z}{L}\left(\s_{-}+z\s_{+}+\cd\right),\\
A_{t} & =\m-z\r+\cd.
\end{align}
In addition, the axial gauge $A_{z}=0$ is adopted in what follows.
According to the dictionary of AdS/CFT duality, $\m$ and $\r$ are
the chemical potential and particle number density of the boundary
system, respectively. To describe superfluidity as a spontaneous U(1)
symmetry breaking, we will switch off the source $\s_{-}$, and then
$\s_{+}$ corresponds to the condensate (expectation value of the
dual operator).

The boundary system is parametrized by $T/\m$. When $T$ is fixed,
a phase transition happens at a critical chemical potential $\m_{c}$,
above which $\s_{+}$ does not vanish and the system is in the superfluid
phase\cite{HHH Model}. We will work in the superfluid phase, so
in the following we will choose $\m>\m_{c}$.

\section{Initial Configurations and Quasi-Normal Modes\label{sec:QNM}}

\subsection{Initial configurations}

We will work in polar coordinates, so the metric of the Schwarzschild-AdS
balck brane (\ref{eq:Sch-AdS}) becomes 
\begin{equation}
ds^{2}=\frac{L^{2}}{z^{2}}\left(-f\left(z\right)dt^{2}+\frac{1}{f\left(z\right)}dz^{2}+dr^{2}+r^{2}d\t^{2}\right).\label{eq:SAdS_polar}
\end{equation}
For numerical simplicity, we set $z_{h}=1$ and $L=1$ from now on.
Hence, $\m_{c}=4.06$ in our numerical computations\cite{HHH Model}.
We will prepare a rotating BEC as our initial state, which is implemented
by taking all fields real and independent of $\t$: 
\begin{equation}
\Psi\left(z,r\right)=\bar{\Psi}\left(z,r\right),\qquad A_{\m}=A_{\m}\left(z,r\right).
\end{equation}
With ansatz given above, we can choose $A_{r}$ to be $0$, which
actually requires no radial current in the bulk\cite{Vortices_JHEP_2010}.
Then the simplified equations of motion are 
\begin{align}
\p_{z}\left(f\p_{z}\s\right)+\p_{r}^{2}\s+\frac{1}{r}\p_{r}\s+\left(\frac{A_{t}^{2}}{f}-\frac{A_{\t}^{2}}{r^{2}}-z\right)\s & =0,\\
f\p_{z}^{2}A_{t}+\p_{r}^{2}A_{t}+\frac{1}{r}\p_{r}A_{t}-2A_{t}\s^{2} & =0,\\
\p_{z}\left(f\p_{z}A_{\t}\right)+\p_{r}^{2}A_{\t}-\frac{1}{r}\p_{r}A_{\t}-2A_{\t}\s^{2} & =0,
\end{align}
where we have made the replacement $\S=z\s$.

At the AdS conformal boundary $z=0$, the boundary conditions 
\begin{equation}
\left.\s\right|_{z=0}=0,\qquad\left.A_{t}\right|_{z=0}=\m,
\end{equation}
are as usual with $\mu$ the chemical potential. For the superfluid
system, a rigid rotation can be introduced by letting 
\begin{equation}
\left.A_{\t}\right|_{z=0}=\Om r^{2},
\end{equation}
where $\Om$ is the angular velocity. The system should have a finite
size in the $r$ direction, i.e. the radius, and in what follows we use
$R$ to denote this radius. The boundary conditions at the edge $r=R$ of the system are taken as 
\begin{equation}
\left.\p_{r}A_{t}\right|_{r=R}=0,\qquad\left.\p_{r}\s\right|_{r=R}=0,\qquad\left.A_{\t}\right|_{r=R}=\Om R^{2}.
\end{equation}
For detailed discussions of vortices in holographic systems, see \cite{Vortices_PRD, Vortices_JHEP_2010}.
In particular, the differences between superfluid vortices and superconductor
vortices in holographic models are discussed in \cite{Vortices_PRD}.
At the center $r=0$ of the polar coordinates, we impose the boundary
conditions 
\begin{equation}
\left.\p_{r}A_{t}\right|_{r=0}=0,\qquad\left.\p_{r}\s\right|_{r=0}=0,\qquad\left.A_{\t}\right|_{r=0}=0,
\end{equation}
from the rotation invariance of the initial configuration.

\begin{figure}
	\centering
	\includegraphics[width=0.44\textwidth]{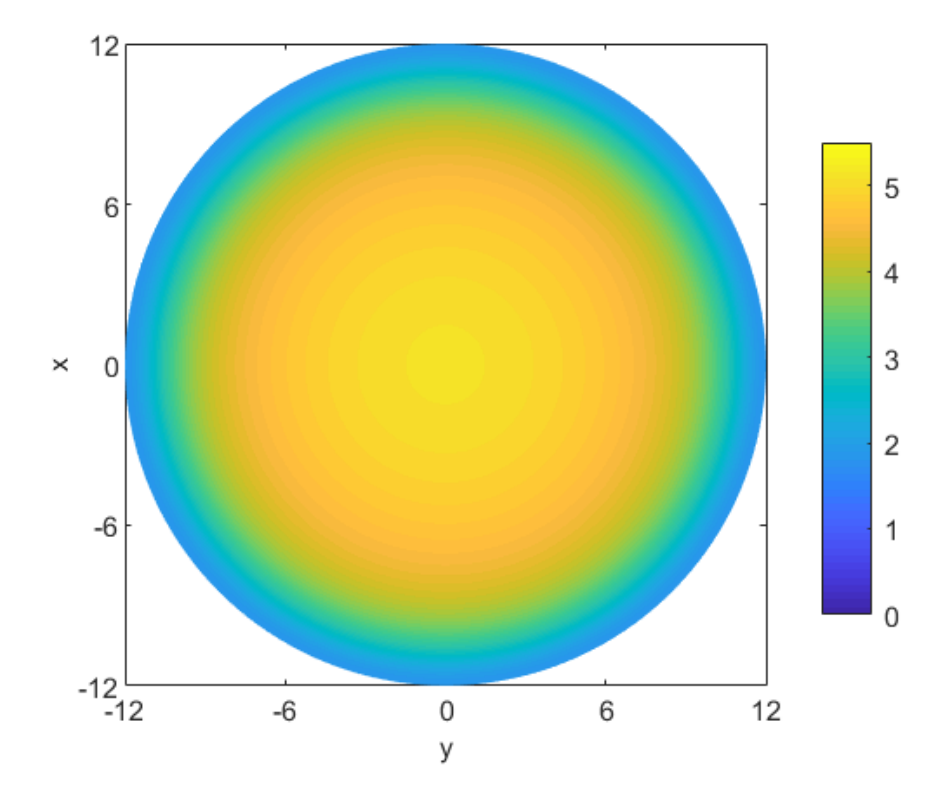}
	\caption{Order parameter as a function of space coordinates when  $\Omega=0.168$.}
	\label{fig:Init-Conf-0.168-boxed}
\end{figure}

\begin{figure}	
	\subfloat[\label{fig:Init-Conf-0.168}Distribution of order parameter when  $\Omega=0.168$.]{\includegraphics[width=0.3\textwidth]{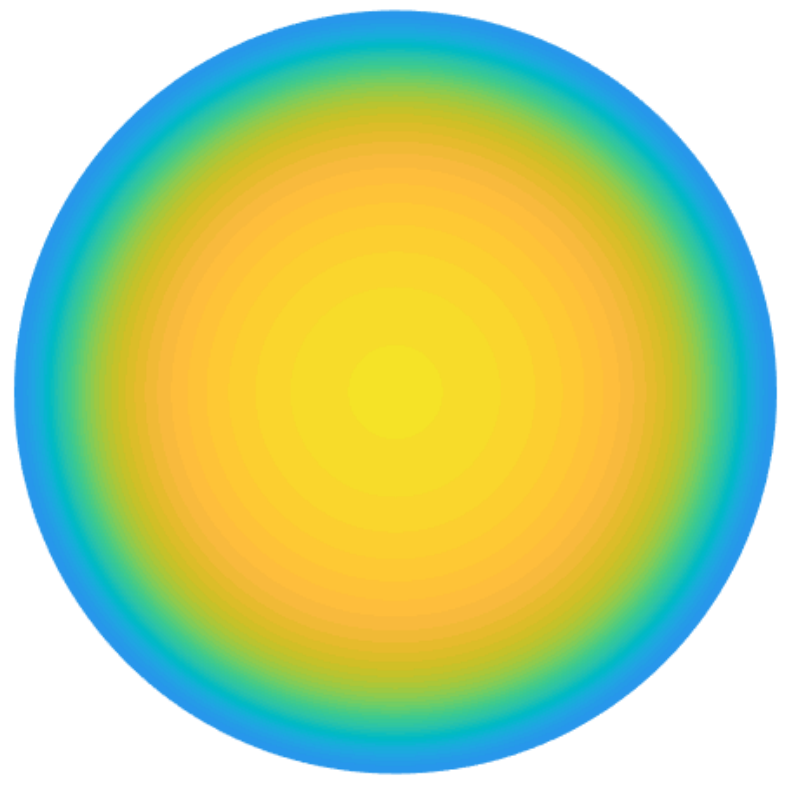}
	}\qquad\qquad\qquad\subfloat[\label{fig:Init-Conf-0.22}Distribution of order parameter when  $\Omega=0.22$.]{\includegraphics[width=0.3\textwidth]{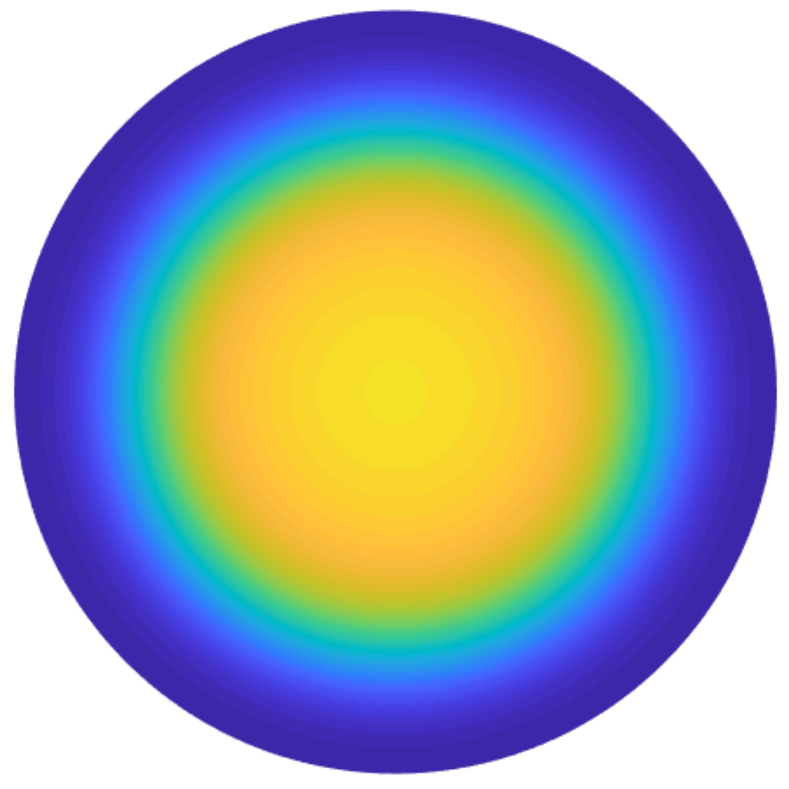}}
	\caption{Initial Configurations with space coordinates ommited.}
	\label{fig:Initial-Configuration}
\end{figure}

From the above equations of motion and boundary conditions, we can obtain the specific initial configuration numerically, once the chemical potential $\mu$ and the angular velocity $\Omega$ are given. For illustration, we plot the distribution of order parameter as a function of space coordinates $(x,y)$ when $\m=5.5$ and $\Om=0.168$ for our initial states in Fig.~\ref{fig:Init-Conf-0.168-boxed}, where we utilise different colors to denote values of order parameter. To make a good comparison, this color scheme will be preserved from now on for all figures of distributions of order parameters in Fig.~\ref{fig:Initial-Configuration}, Fig.~\ref{fig:Lattice-7}, Fig.~\ref{fig:Lattice-19} and Fig.~\ref{fig:Lattice-18}. From now on, we choose $R=12$ in all our numerical studies except for Fig.~\ref{fig:CAV_R}, Fig.~\ref{fig:CLV_R}, Fig.~\ref{fig:Num_Vortices} and Fig.~\ref{fig:Equlibriation_Time}.

For simplicity, we will omit space coordinates axes when plotting distributions of order parameters in all following plots for order parameters. We replot the distribution of order parameter in Fig.~\ref{fig:Init-Conf-0.168} when $\Om=0.168$ with coordinates ommited, and Fig.~\ref{fig:Init-Conf-0.22} shows distribution of order parameter when $\Om=0.22$. Here, the distribution in Fig.~\ref{fig:Init-Conf-0.168} will be used as initial state for generation of lattice with 7 vortices, and the distribution in Fig.~\ref{fig:Init-Conf-0.22} is initial state for generation of lattices with 18 and 19 vortices. Fig.~\ref{fig:Initial-Configuration} shows that with the increasement of $\Om$, boundary values of order parameters are depressed, and this is the expected result when the rotating superfluid consists of no vortex.

\subsection{Quasi-normal modes \label{subsec:Quasi-normal-modes}}

Before performing the nonlinear real time evolution, we first make a linear analysis of the system, which corresponds to quasi-normal modes of the bulk
system. From now on, we will switch to the ingoing Eddington-Finkelstein
coordinates, so that the ingoing boundary conditions at the horizon
are easily incorporated by regularity. Under the Eddington-Finkelstein
coordinates, the metric (\ref{eq:SAdS_polar}) becomes 
\begin{equation}
ds^{2}=\frac{1}{z^{2}}\left(-f\left(z\right)dt^{2}-2dtdz+dr^{2}+r^{2}d\t^{2}\right).\label{eq:metric}
\end{equation}
Note that the original axial gauge $A_{z}=0$ in the Schwarzschild
coordinates (\ref{eq:SAdS_polar}) is violated after this coordinate
transformation. In order to preserve $A_{z}=0$ in the new coordinates,
a U(1) gauge transformation should be performed after the coordinate
transformation: 
\begin{equation}
\s\left(z,r\right)\to e^{i\l\left(z,r\right)}\s\left(z,r\right),\qquad\l=-\i\frac{A_{t}}{f}dz.
\end{equation}
As a result, $A_{r}=\p_{r}\l\left(z,r\right)$ and so it no longer
vanishes, while $\s$ should not be taken real anymore.

Since the background configurations constructed in the last subsection
are time translation invariant and rotation invariant, the linear
perturbations can be expanded as\cite{DLTZ} 
\begin{align}
\d\s & =p\left(z,r\right)e^{-i\o t+im\t}+q^{*}\left(z,r\right)e^{i\o^{*}t-im\t}, \label{eq:perturbation_of_psi}\\
\d A_{\m} & =a_{\m}\left(z,r\right)e^{-i\o t+im\t}+a_{\m}^{*}\left(z,r\right)e^{i\o^{*}t-im\t},\label{eq:perturbation_of_A}
\end{align}
where the coefficients $p$, $q$ and $a_{\mu}$ are all complex functions
of $(z,r)$. In particular, $\d\s$ is complex, so $q\left(z,r\right)$
is independent of $p\left(z,r\right)$. Equations for quasi-normal modes are shown in Appendix \ref{sec:EOMs_for_QNMs}. 

At the AdS conformal boundary $z=0$, Dirichlet boundary conditions
should be imposed \cite{DLTZ}: 
\begin{equation}
\left.p\right|_{z=0}=0,\;\left.q\right|_{z=0}=0,\;\left.a_{t}\right|_{z=0}=0,\;\left.a_{r}\right|_{z=0}=0,\;\left.a_{\t}\right|_{z=0}=0.\label{eq:BC_conform}
\end{equation}
At $r=0$, boundary conditions can be obtained by asymptotic analyses
of the perturbation equations: 
\begin{equation}
\left.\p_{r}p\right|_{r=0}=0,\;\left.\p_{r}q\right|_{r=0}=0,\;\left.\p_{r}a_{t}\right|_{r=0}=0,\;\left.a_{r}\right|_{r=0}=0,\;\left.\p_{r}a_{\t}\right|_{r=0}=0.\label{eq:BC_polar}
\end{equation}
Boundary conditions at $r=R$ should be consistent with static case:
\begin{equation}
\left.\partial_{r}p\right|_{r=R}=0,\;
\left.\partial_{r}q\right|_{r=R}=0,\;
\left.\partial_{r}a_t\right|_{r=R}=0,\;
\left.a_{r}\right|_{r=R}=0,\; 
\left.a_{\theta}\right|_{r=R}=0.
\end{equation}

\begin{figure}
	\subfloat[\label{fig:Stability-of-each}The most unstable modes as a function
	of $m$ when $\protect\Om=0.168$ and $\protect\m=0.55$. In this plot,
	the instability reaches its maximum when $m=19$.]{\includegraphics[width=0.32\textwidth]{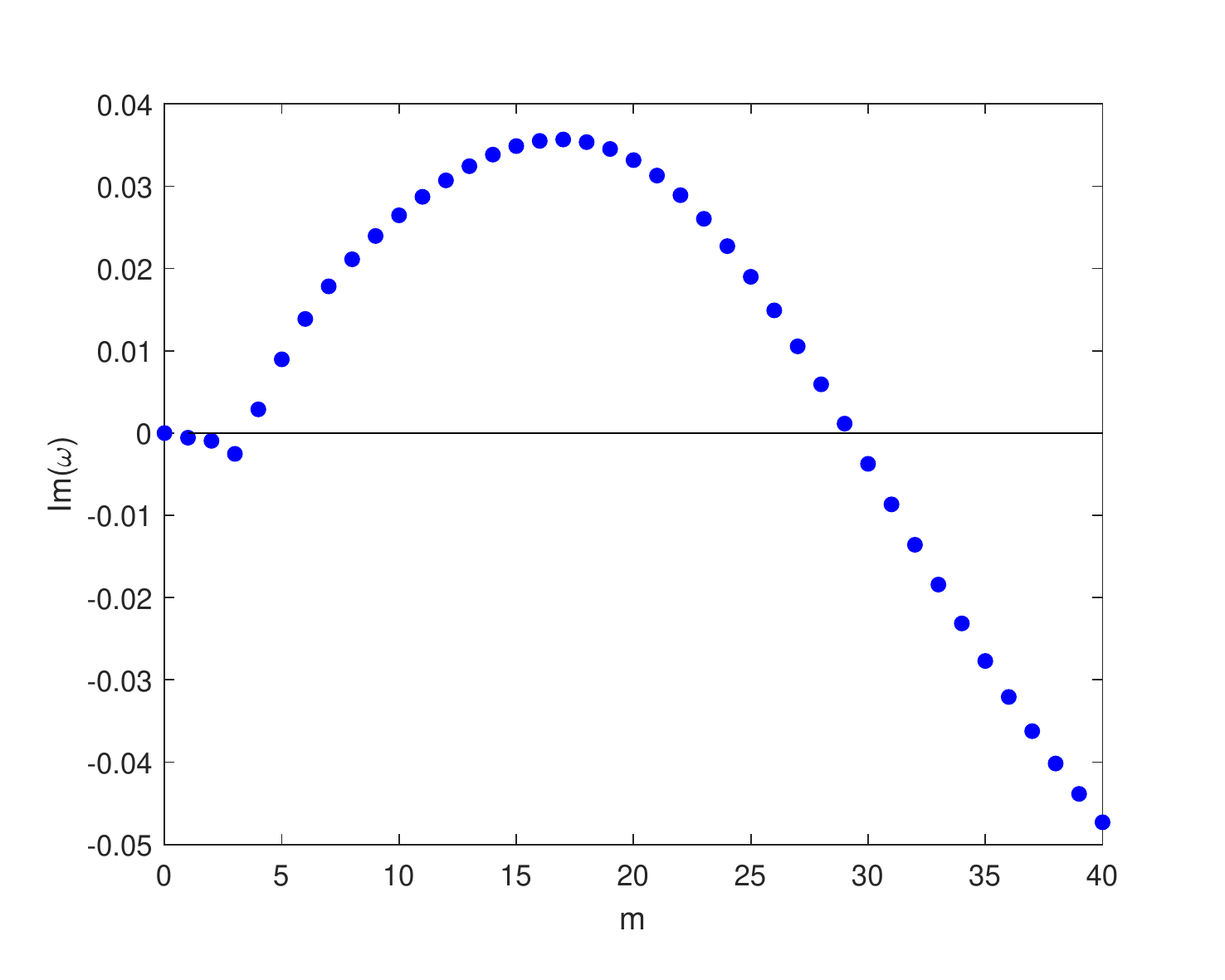}
	}\subfloat[\label{fig:Critical-Angular-Velocity}Critical angular velocities
	calculated with different values of the parameter $m$. In this plot, the critical velocities reach their minimum when $m=15$.]{\includegraphics[width=0.32\textwidth]{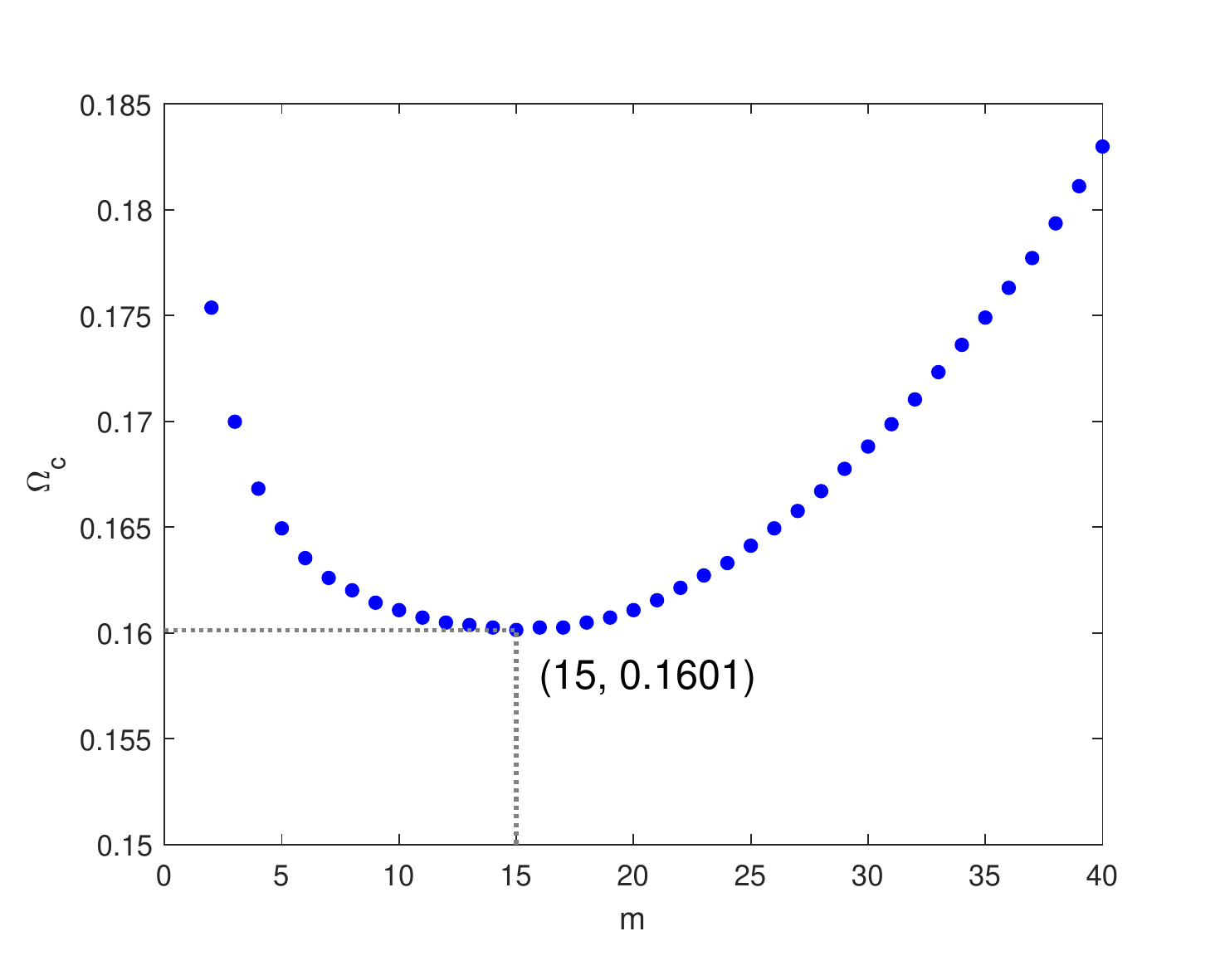}		
	}\subfloat[\label{fig:CAV_mu}Critical angular velocity
	as a function of $\mu$. In this plot, the critical velocity grows with the increase of $\mu$.]{\includegraphics[width=0.32\textwidth]{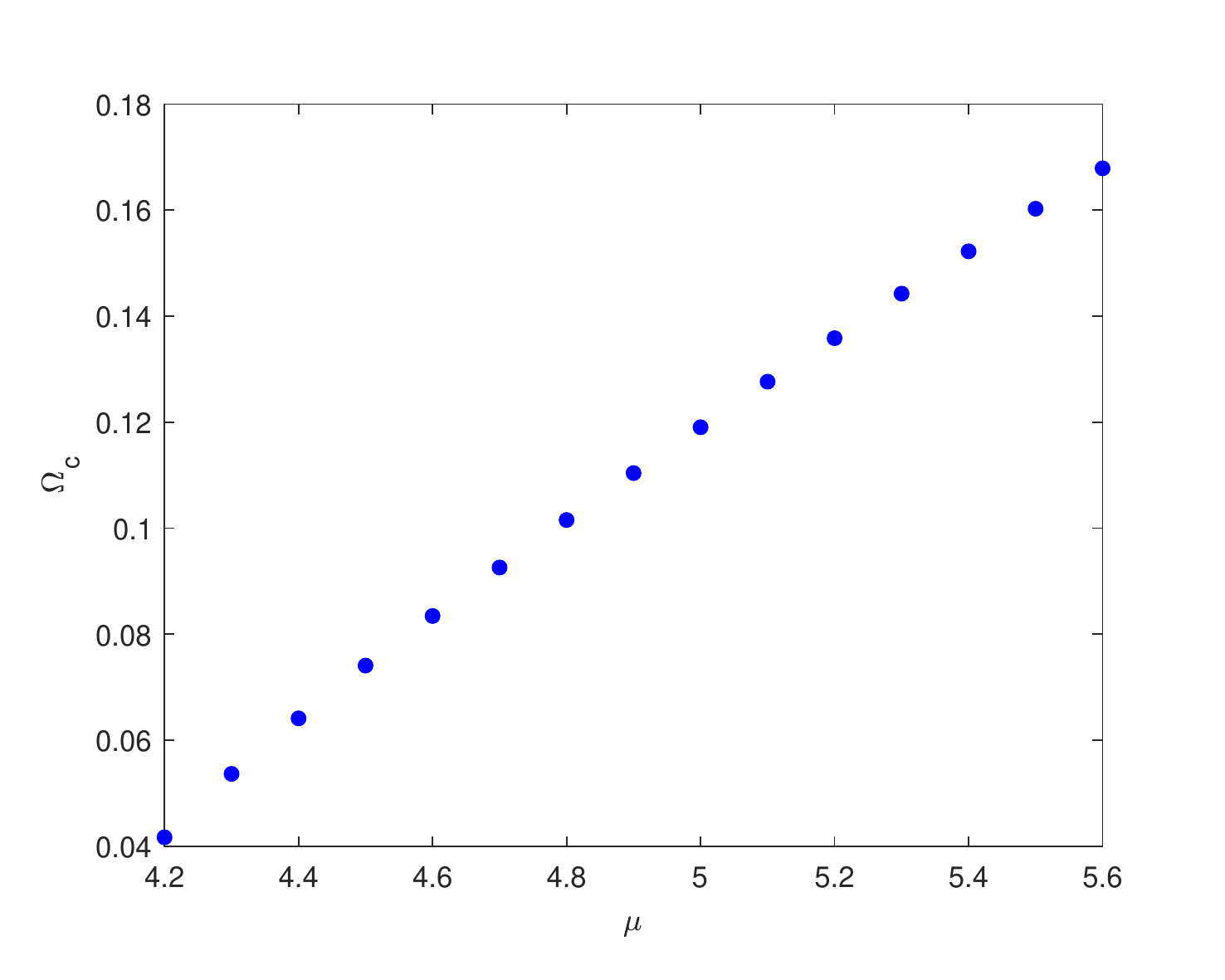}		
	}\qquad\flushleft\subfloat[\label{fig:CAV_R}Critical angular velocity
	as a function of $R$. Critical angular velocity decreases with the  increase of $R$.]{\includegraphics[width=0.32\textwidth]{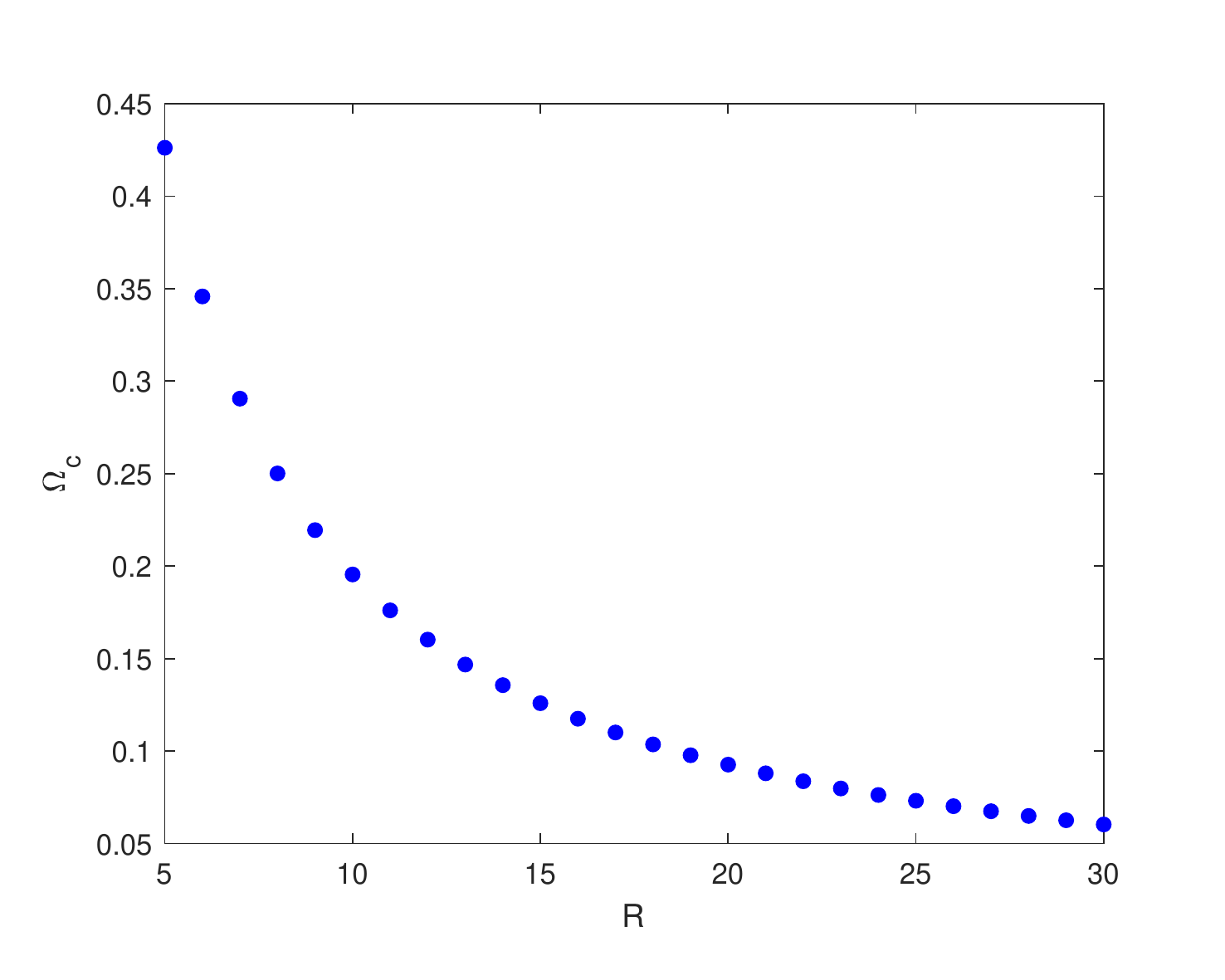}		
	}\subfloat[\label{fig:CLV_R}Linear velocity $v|_{r=R}$ as a function of $R$. In this plot, $\Delta R=25$ while $\Delta v|_{r=R}\approx0.3$.]{\includegraphics[width=0.32\textwidth]{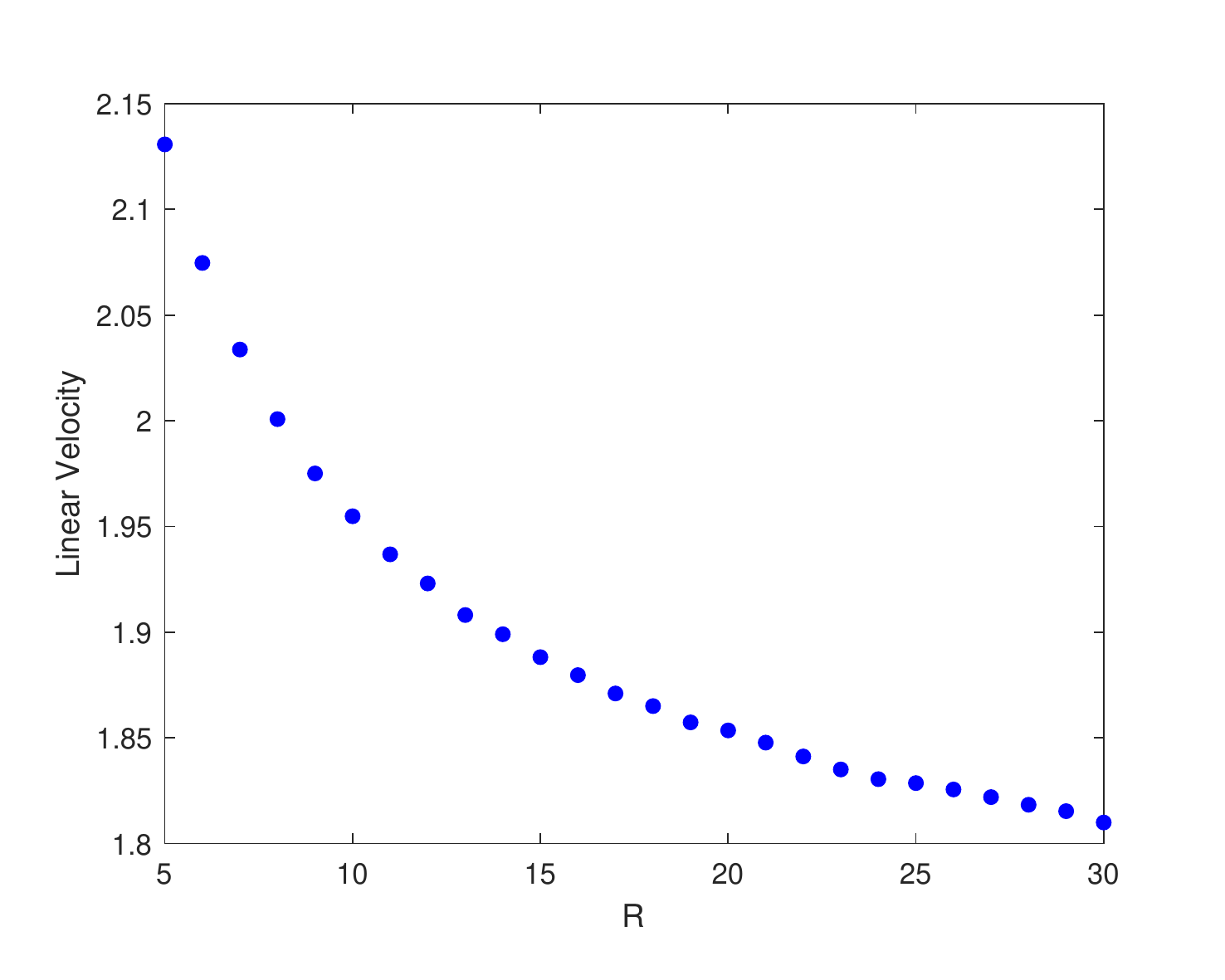}		
}\caption{Quasi-Normal Modes}
\end{figure}

On top of a background configuration constructed in the last subsection and for a given $m$, the quasi-normal frequencies $\omega$ are determined
by the condition that the perturbation equations together with the
boundary conditions (\ref{eq:BC_conform}) and (\ref{eq:BC_polar})
have nontrivial solutions. Actually, for every $m$, we can obain
a series of $\o$, among which the one with the maximal imaginary
part determines the instability (if there is any $\omega$ with a
positive imaginary part) of the $m$ mode: this mode is more unstable
if the imaginary part of this $\o$ is larger. Fig.\ref{fig:Stability-of-each}
shows our results for the background configuration in Fig.~\ref{fig:Init-Conf-0.168} with $\Om=0.168$,
where we can see that the system is unstable against the perturbation
of the $4\leq m\leq29$ modes and the most unstable mode has $m=19$ \footnote{Instability of modes when $\Om=0.22$ is similar.}.

A critical angular velocity $\Omega_{m}$ can be determined for every $m$, which is just the lowest value of $\Omega$ that makes the $m$ mode unstable. The critical angular velocities for different $m$ are shown in Fig.\ref{fig:Critical-Angular-Velocity}. We can see that the critical angular velocity for the system is $\Omega_{c}\approx0.16$, which is the lowest value among all $\Omega_{m}$s. If the angular velocity $\Omega<\Omega_{c}$, the rotating superfluid system is stable and there will be no vortex generation. In addition, relations between $\Omega_{c}$ and chemical potential $\mu$ as well as $R$ are studied. In Fig.~\ref{fig:CAV_mu}, the critical angular velocity $\Omega_{c}$ is calculated as a function of $\mu$. Our result shows that $\Omega_{c}$ grows with $\mu$, which indicates that themal fluctuations contribute to instability of the system. In Fig.~\ref{fig:CAV_R}, the critical angular velocity $\Omega_{c}$ is calculated with $R$ varying from $5$ to $30$, which shows that the critical angular velocity $\Omega_{c}$ decreases with the rise of $R$. In order to understand this relation better, we calculate the linear velocity $v$ at $r=R$ in Fig.~\ref{fig:CLV_R}. We find that compared with the scale of the system, the change of linear velocity $\Delta v$ is rather small. This rather small $\Delta v$, as well as the result from nonlinear evolutions in the next section that the most obvious change of the system first appears at the edge, indicates that the instability of the system should be mainly determined by the linear velocity at the edge $r=R$.

\section{Nonlinear Time Evolution\label{sec:Nonlinear}}

\subsection{Formation of vortex lattices}

When the angular velocity $\Om$ is above the critical value $\Omega_{c}$,
the system is unstable. In real systems, there are random initial
perturbations from environment noises, thermal and quantum fluctuations,
imperfect preparation of the initial configurations and so on. If
the system is unstable, the tiny perturbations of the unstable modes
will exponentially grow up at the early stage and then go beyond the
linear regime. Will the instability when $\Omega>\Omega_{c}$ finally
result in vortex generation and vortex lattice formation? To answer
this question, which is very important but outside the scope of the
linear analyses, we will slightly perturb the system from the initial
configurations constructed in Sec.~\ref{sec:QNM}, and see what will
happen by numerically simulating the nonlinear dynamics of this holographic
system. To be specific, we will set $\m=5.5$ from now on.

Basically, we use the same strategy as in \cite{DNTZ,LTZ,GKLTZ}
for the numerical evolution here. The only difference and complication
is the polar coordinates for this system, which has never been coped
with in nonlinear time evolution of holographic superfluids. The full
equations of motion in the polar coordinates are shown in Appendix \ref{sec:EOMs_for_Evolution}.

At every step of this evolution, we solve $A_{t}$ from the restriction
equation (\ref{eq:restrict}) and use (\ref{eq:evolve_psi},\ref{eq:evolve_Ar},\ref{eq:evolve_Ah})
to evolve $(\psi,A_{r},A_{\theta})$, while equation (\ref{eq:boudary_rho})
is used to set a boundary condition for $A_{t}$ (recalling $\r=\left.-\p_{z}A_{t}\right|_{z=0}$):
\begin{equation}
\p_{t}\r=\left.-\left(\frac{1}{r}\p_{z}A_{r}+\p_{z}\p_{r}A_{r}+\frac{1}{r^{2}}\p_{z}\p_{\t}A_{\t}\right)\right|_{z=0}.
\end{equation}
We also impose the boundary conditions at $z=0$ as 
\begin{equation}
\left.\p_{t}\s\right|_{z=0}=0,\;\left.A_{t}\right|_{z=0}=\m,\ \left.\p_{t}A_{r}\right|_{z=0}=0,\ \left.\p_{t}A_{\t}\right|_{z=0}=0.
\end{equation}
At the edge $r=R$ of the system, we set 
\begin{equation}
\left.\p_{t}\p_{r}\s\right|_{r=R}=0,\ \left.\p_{t}A_{r}\right|_{r=R}=0,\ \left.\p_{t}A_{\t}\right|_{r=R}=0.
\end{equation}
We do spectral expansion in the $(z,r,\theta)$ directions, but skip
the coordinate singularity at $r=0$, so no boundary conditions are
needed there\cite{Spectral Methods}. As well, under the ingoing
Eddington-Finkelstein coordinates, no boundary conditions are needed at the horizon.

To evolve the system, initial perturbations are necessary. In our time evolution, perturbations at $t=0$ are imposed merely on $\psi$ with the following form:
\begin{equation}
	\delta\psi=\sum_{m} A_m \cdot z \sin^m\left(\frac{\pi r}{2R}\right)\exp(im\theta),
	\label{eq:perturbation}
\end{equation}
Here, $A_m$ is the amplitude which should be small enough, and $z$ preserves Dirichlet boundary condition for $\psi$. $\sin^m\left(\frac{\pi r}{2R}\right)$ behaves as $r^m$ when $r\to 0$, which is required by the analytic property of $\psi$, and preserves the Neumann boundary condition of $\psi$ at $r=R$.

\begin{figure}
	\subfloat[\label{fig:t=00003D00003D30}$t=30$]{\includegraphics[width=0.2\textwidth]{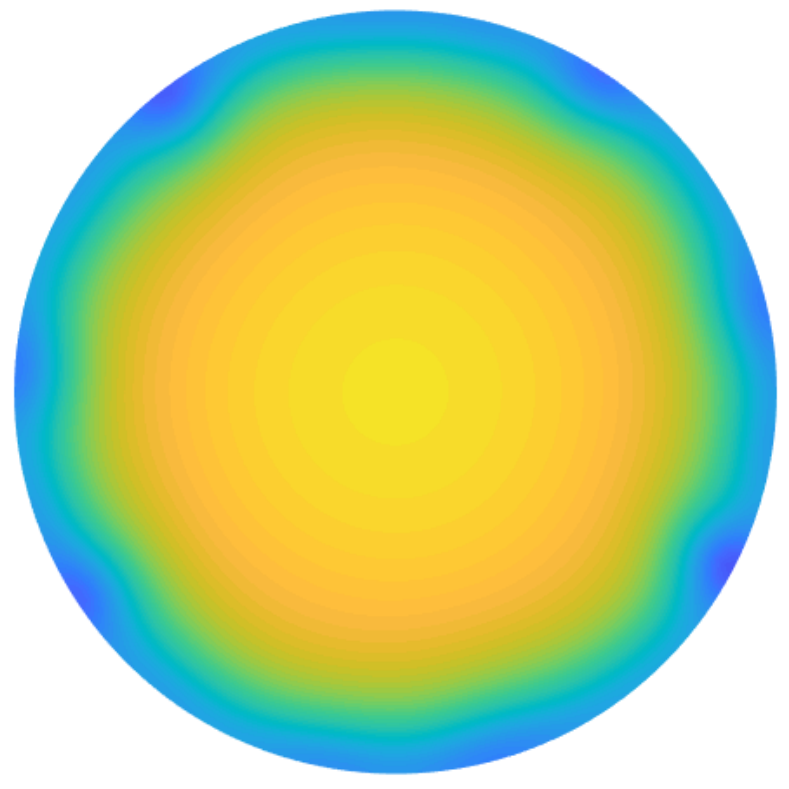}		
	}\subfloat[\label{fig:t=00003D00003D60}$t=60$]{\includegraphics[width=0.2\textwidth]{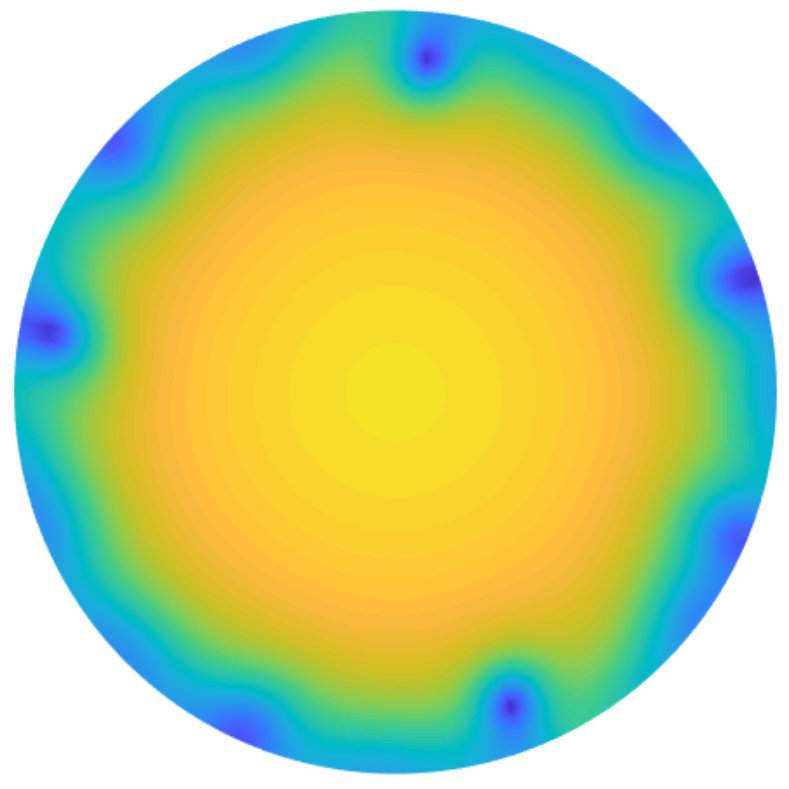}		
	}\subfloat[\label{fig:t=00003D00003D120}$t=120$]{\includegraphics[width=0.2\textwidth]{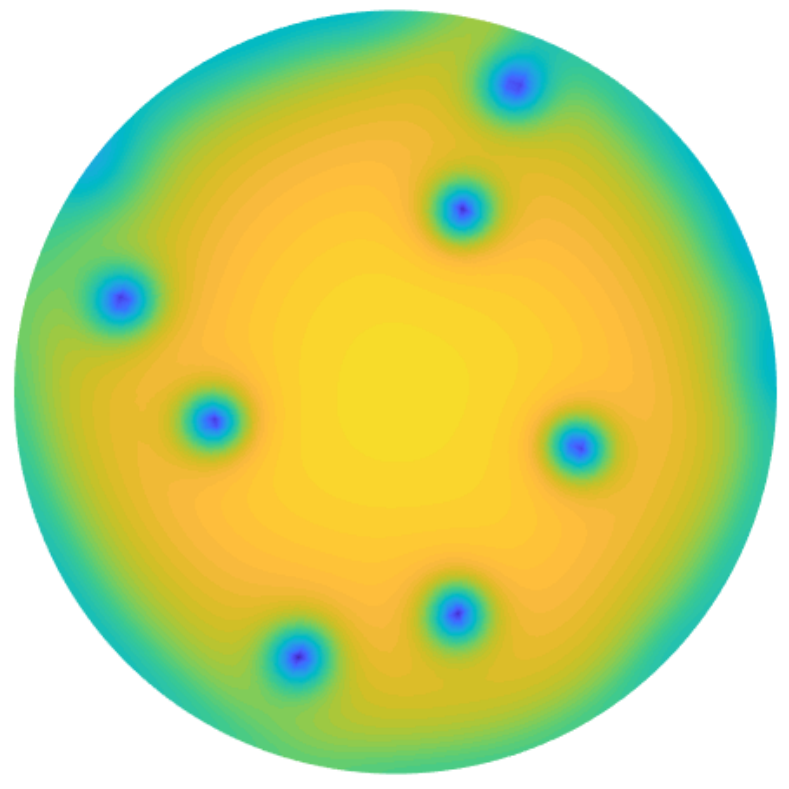}		
	}\subfloat[\label{fig:t=00003D00003D360}$t=360$]{\includegraphics[width=0.2\textwidth]{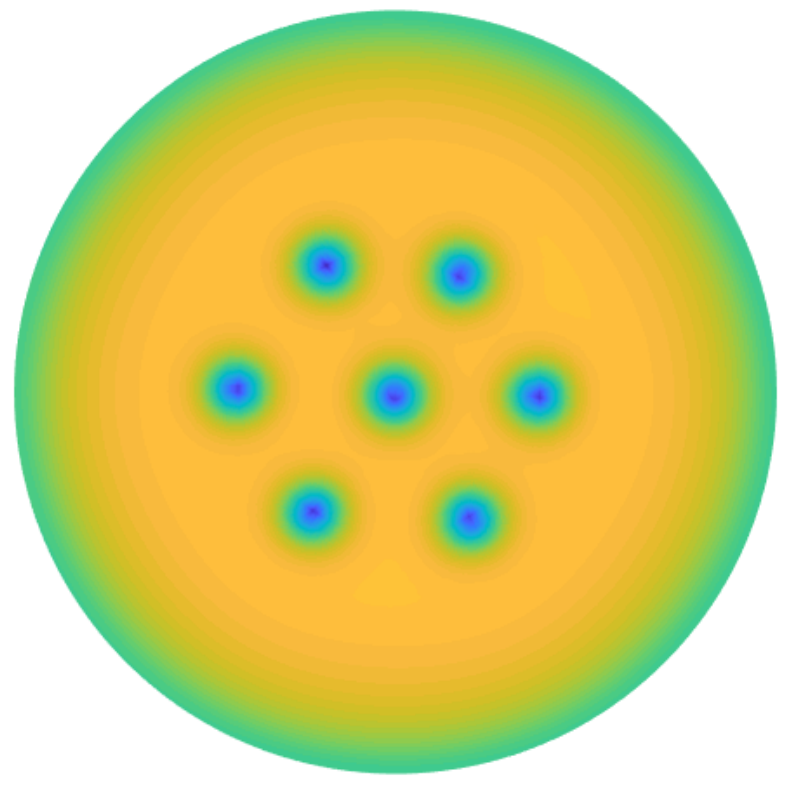}		
	}\subfloat[\label{fig:t=00003D00003D500}$t=500$]{\includegraphics[width=0.2\textwidth]{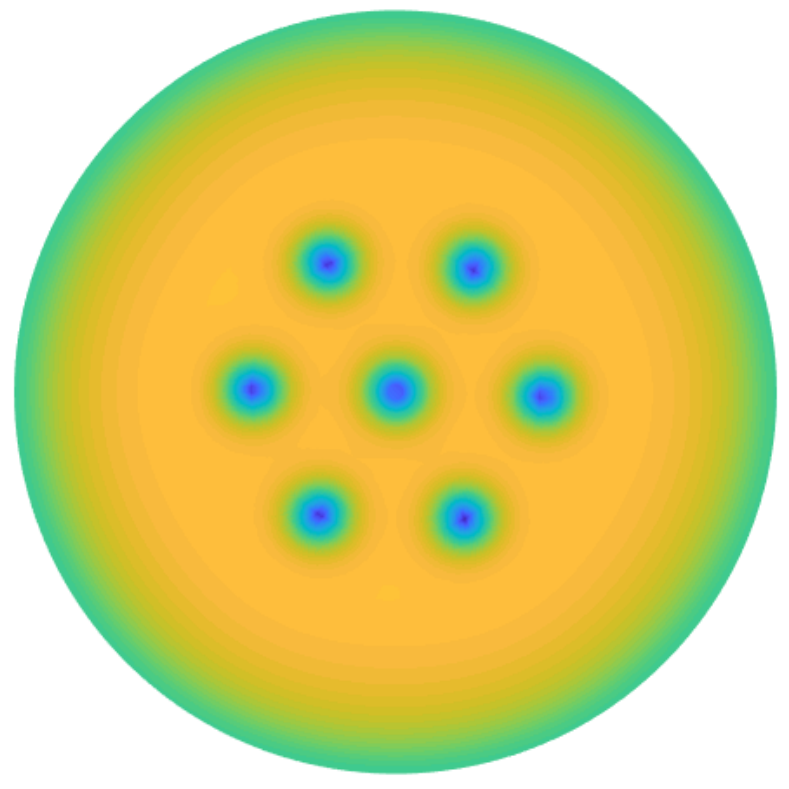}		
	}\caption{\label{fig:Lattice-7}Condensates as functions of space coordinates at different time when $\Omega=0.168$. In this plot, 7 vortices appear, which form the simplest triangular lattice.}
\end{figure}

With the setup above, we can do the time evolution by the standard
fourth order Runge-Kuta method. Here, we will mainly show our numerical results
for a typical process of the vortex lattice formation when $\Om=0.168$.
The results are as follows: When $t<60$, some distortions appear
at the edge of the superfluid. These distortions come from instabilities
we studied in Sec.~\ref{subsec:Quasi-normal-modes} and will lead
to the generation of vortices (see for example at $t=30$ in Fig.~\ref{fig:t=00003D00003D30}). When $t\approx60$, vortices appear
at the edge (Fig.~\ref{fig:t=00003D00003D60}). These vortices then
move toward the center of the superfluid with the ongoing generation
of new vortices at the edge. When $t\approx120$, the generation process
ends, while the movement of vortices continues (Fig.~\ref{fig:t=00003D00003D120}).
Vortices' movement basically ends when $t\approx360$, at which the
lattice forms (Fig.~\ref{fig:t=00003D00003D360}). We continue the
time evolution and find that the structure of the lattice does not
change anymore (see Fig.~\ref{fig:t=00003D00003D500}), which means
that the lattice gets stabilized.

\begin{figure}
	\subfloat[$t=16$]{\includegraphics[width=0.2\textwidth]{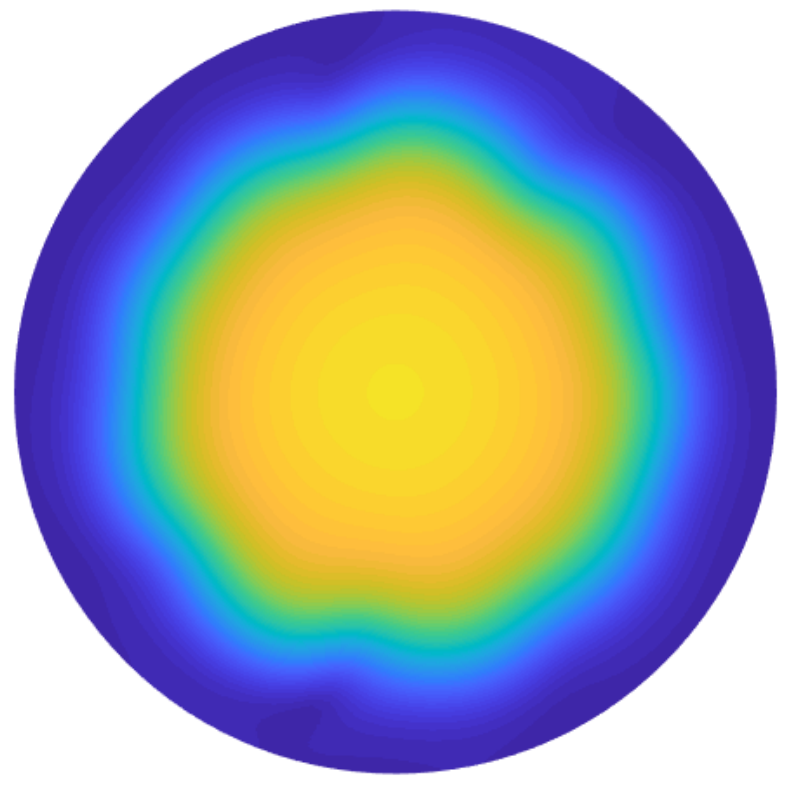}
	}\subfloat[$t=32$]{\includegraphics[width=0.2\textwidth]{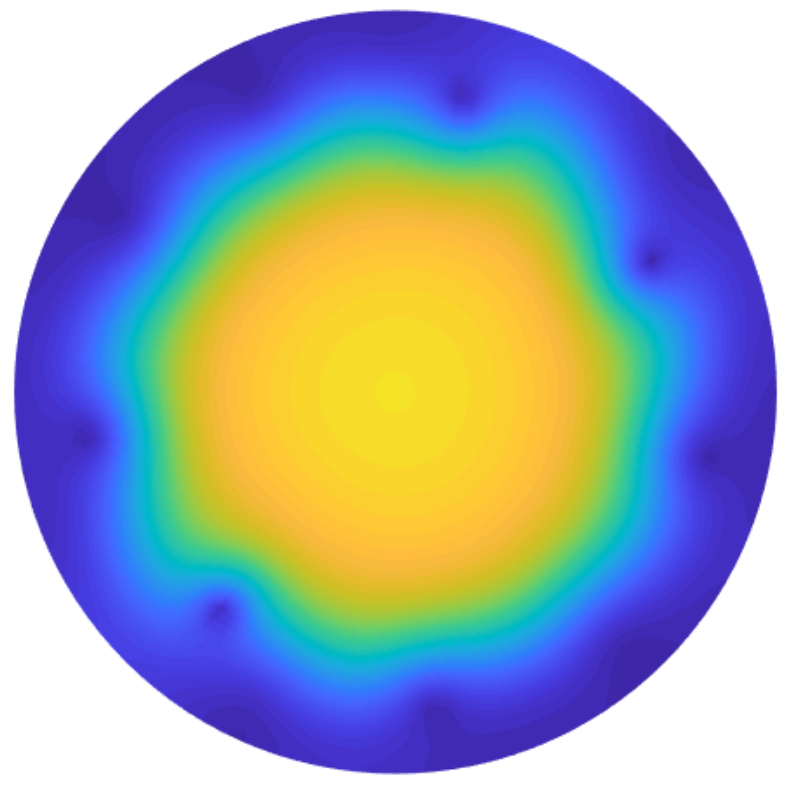}
	}\subfloat[$t=120$]{\includegraphics[width=0.2\textwidth]{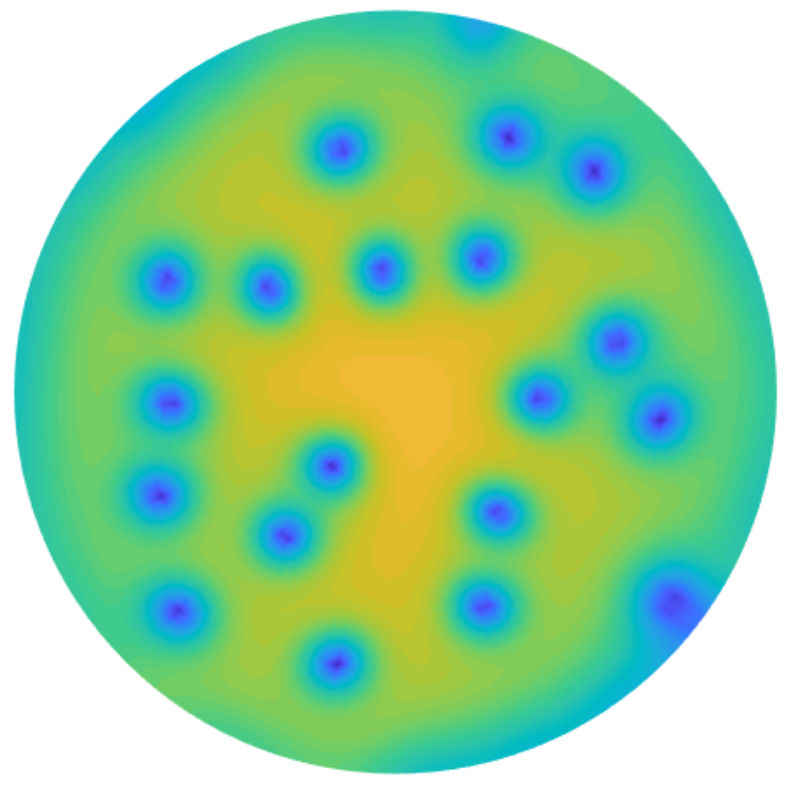}
	}\subfloat[$t=320$]{\includegraphics[width=0.2\textwidth]{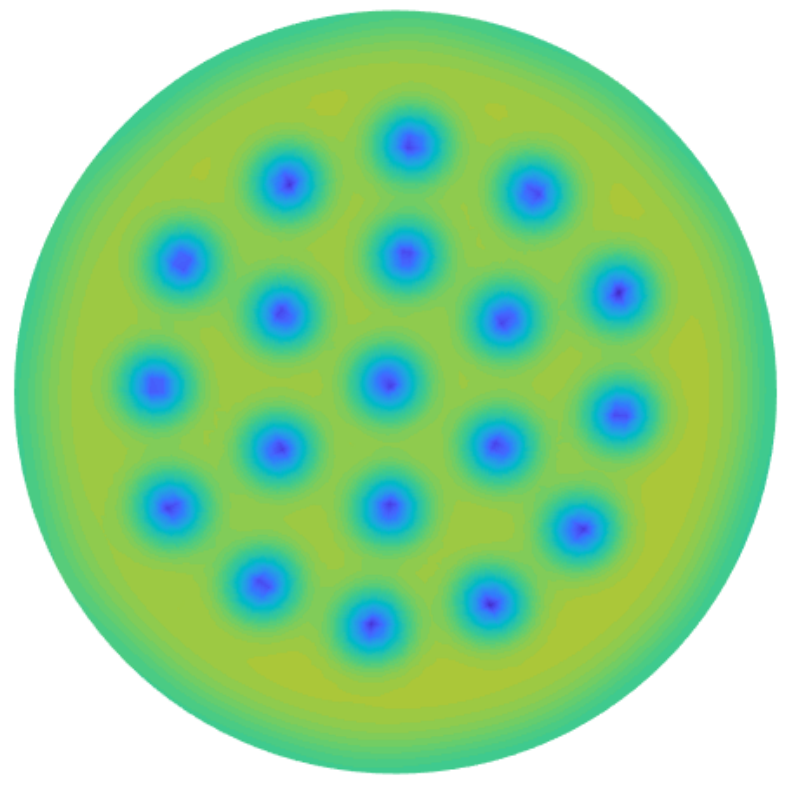}
	}\subfloat[$t=500$\label{fig:Lattice-19-500}]{\includegraphics[width=0.2\textwidth]{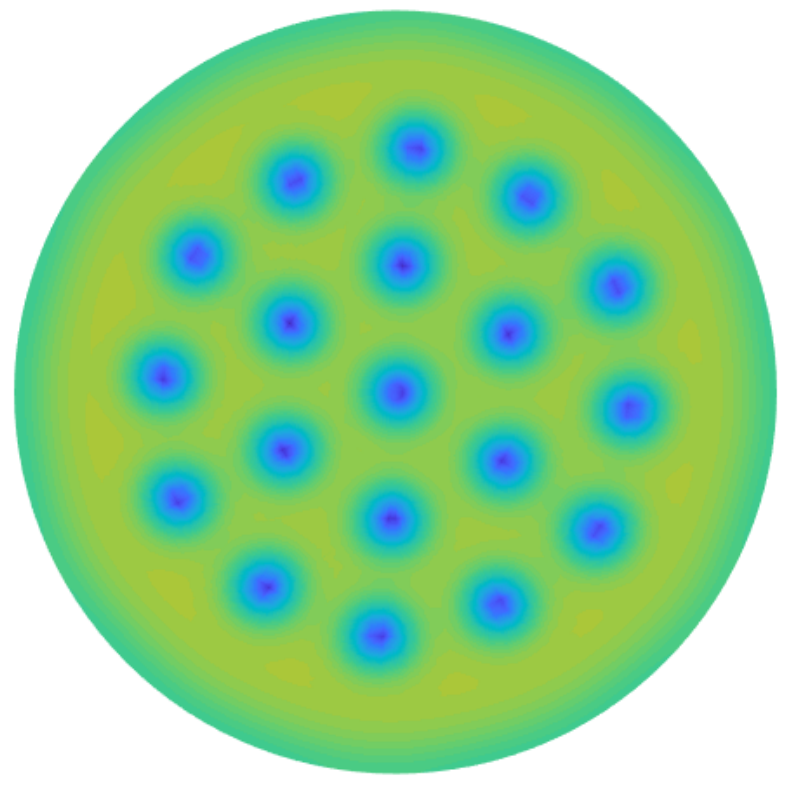}}
	\caption{Condensates as functions of space coordinates when the lattice consists $19$ vortices. In this plot, we choose $\Om=0.22$.}
	\label{fig:Lattice-19}
\end{figure}

To make a good comparison, we show another typical evolution that consists of 19 vortices in Fig.~\ref{fig:Lattice-19}.\footnote{Movies for the time evolution are available at {[}http://people.ucas.edu.cn/\textasciitilde ytian?language=en\#\%20687316{]}.} The evolution process is similar, except for the final background order parameters. In both cases, vortices are arranged triangularly. A more complex lattice will be obtained if we study the case with more vortices in a larger system.

\begin{figure}
	\subfloat[Relations between the number of vortices and $\Omega$. In this plot, we fix $\mu=5.5$, $R=18$ and initial perturbations.\label{fig:Num_Vortices_Omega}]{\includegraphics[width=0.4\textwidth]{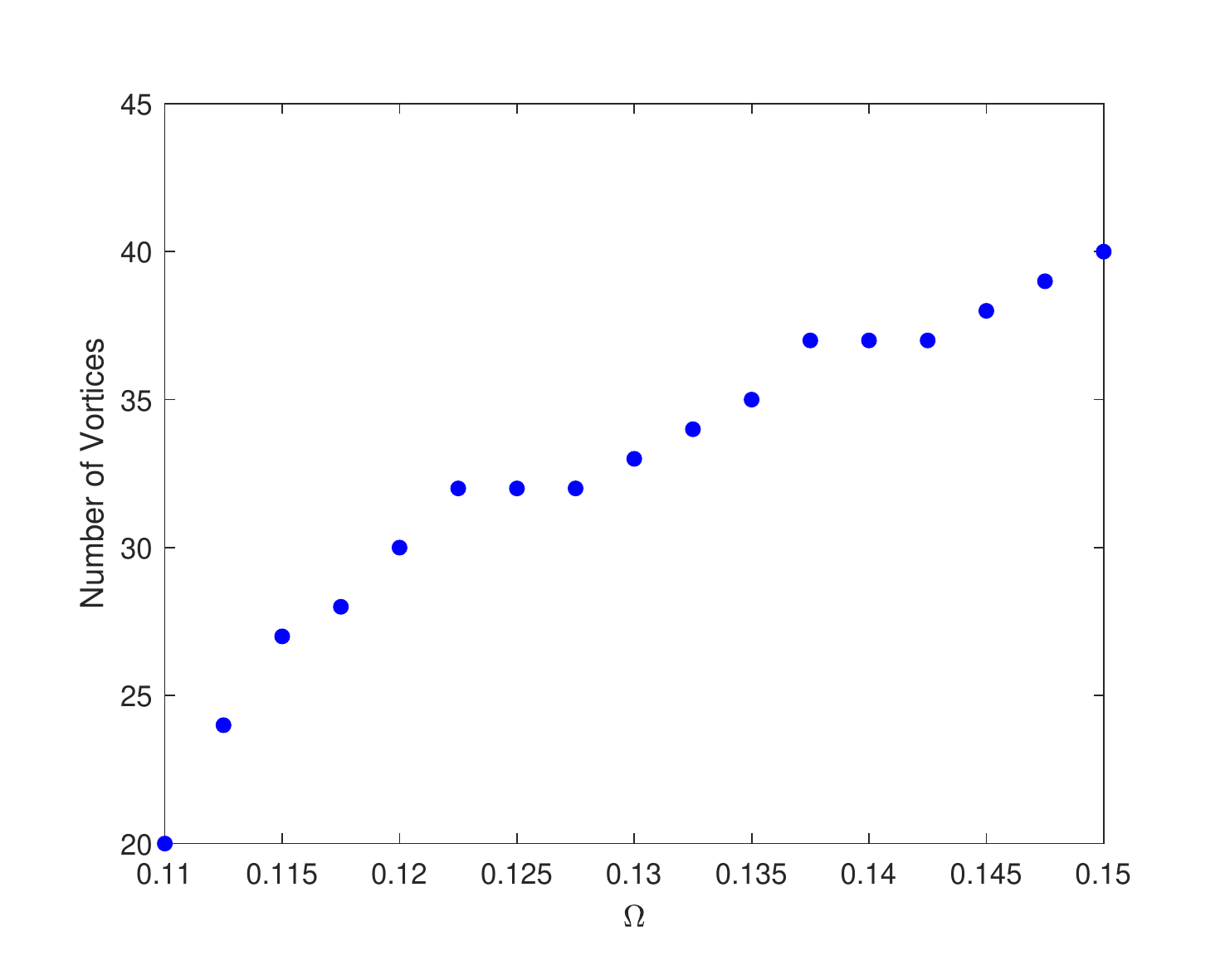}}
	\qquad
	\subfloat[Relations between the number of vortices and $\mu$. In this plot, we fix $\Omega=0.11$, $R=18$ and initial perturbations.\label{fig:Num_Vortices_mu}]{\includegraphics[width=0.4\textwidth]{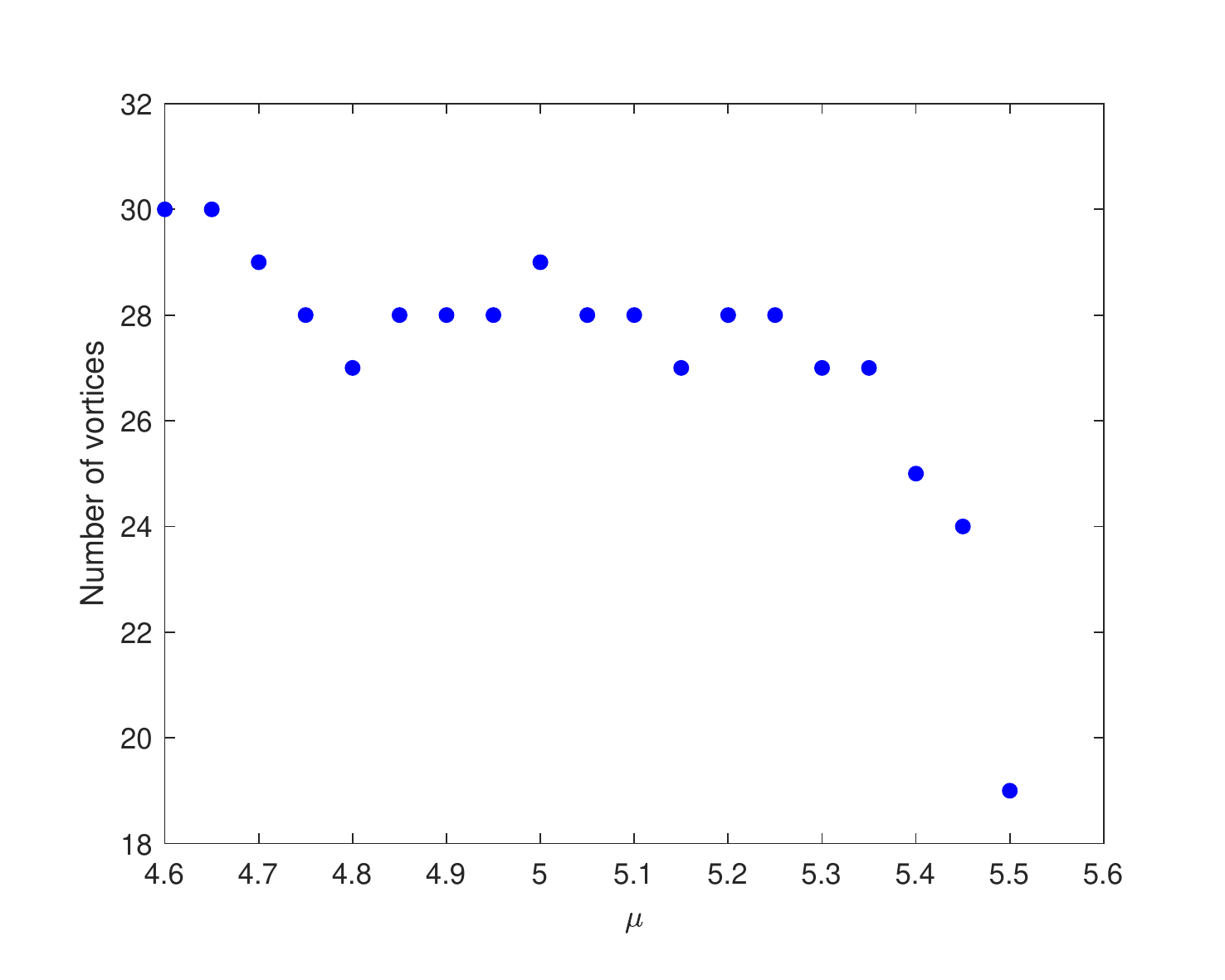}}	
	\caption{Dependence of the vortex number on other quantities.}\label{fig:Num_Vortices}
\end{figure}

It is important to study the relation between the number of vortices and other conditions or parameters, such as the angular velocity and chemical potential, of the system. In order to mimic best the experimental situations, we choose our initial perturbations as equation (\ref{eq:perturbation}) with all unstable modes of the same (tiny enough) amplitude. For the relation between the vortex number and the angular velocity, we fix $\mu=5.5$, $R=18$ and perform evolutions. The result is shown in Fig.~\ref{fig:Num_Vortices_Omega}, where an approximately linear increase of vortex number is witnessed when $\Omega$ grows. This relation is consistent with the Feynman rules\cite{R-books}. Moreover, the relation between the vortex number and $\mu$ is plotted in Fig.~\ref{fig:Num_Vortices_mu}, with fixed $\Omega=0.11$ and $R=18$. We find that there's a tendency that the vortex number decreases with the increase of $\mu$. Overall, a larger $\Omega$ and smaller $\mu$ can generally increase the number of vortices.

\subsection{Properties of Dissipation}

\begin{figure}
	\subfloat[\label{fig:Dissipation-7}Energy dissipation as a function of time when the lattice consists of 7 vortices.]{\includegraphics[width=0.4\textwidth]{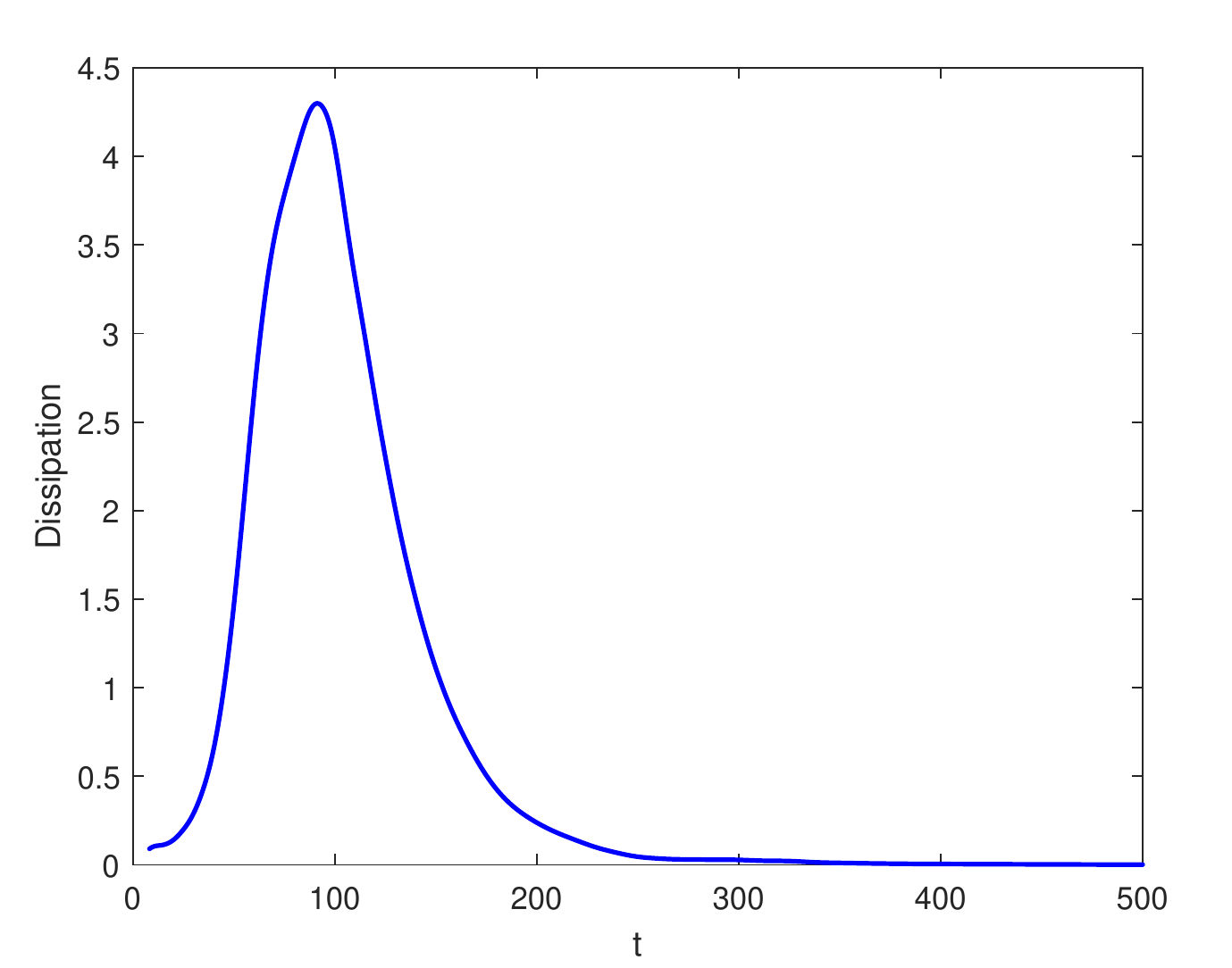}}
	\qquad
	\subfloat[\label{fig:Dissipation-19}Energy dissipation as a function of time when the lattice consists of 19 vortices.]{\includegraphics[width=0.4\textwidth]{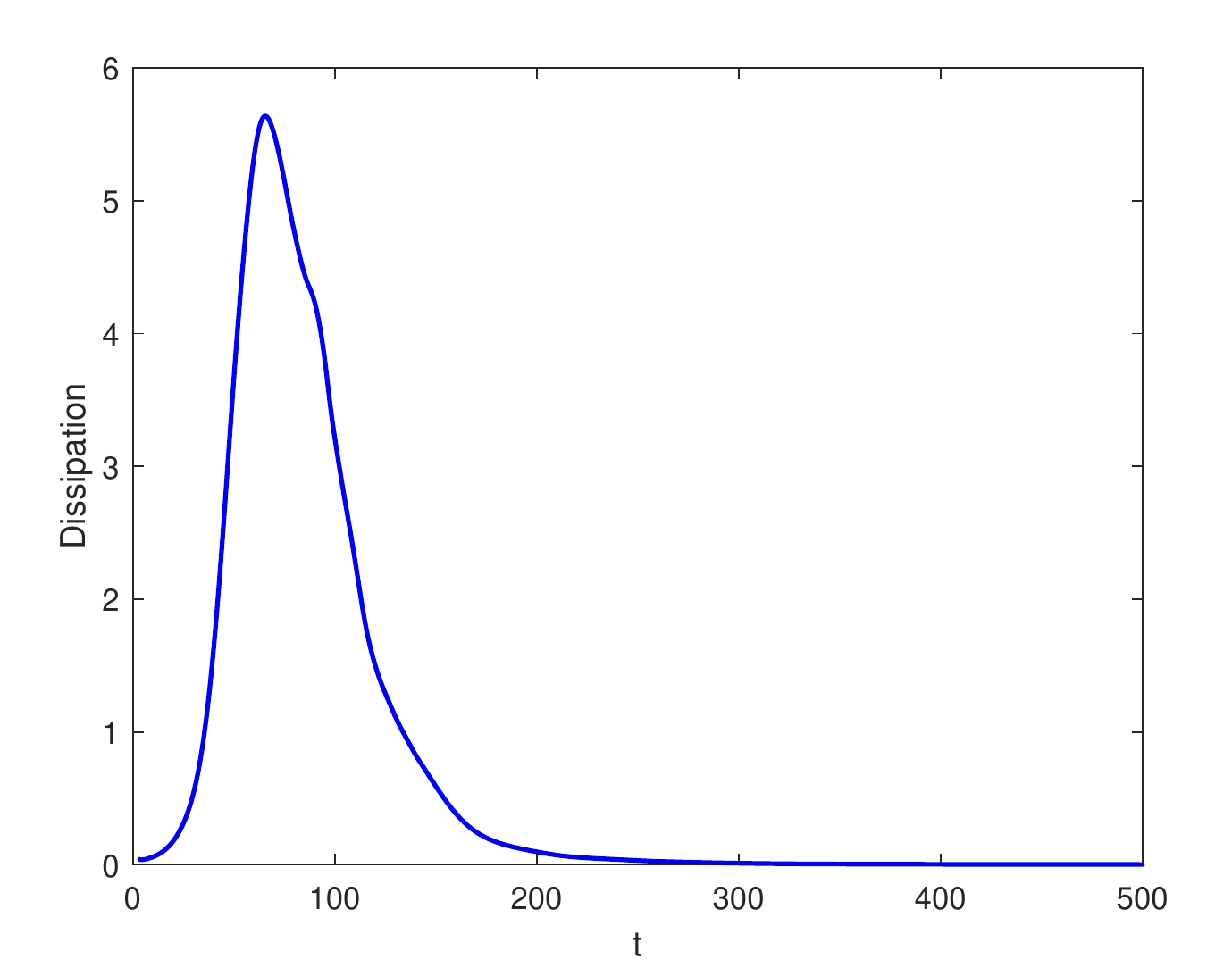}}
	\caption{Energy dissipation when $R=12$.}\label{fig:Dissipations}
\end{figure}

To make a vortex lattice eventually stabilized,	energy dissipation is necessary. In holography, the energy dissipated in this process corresponds to the energy falling into the black hole\cite{Dissipation and Energy Flow,TWZ,Dissipation}. To be precise, the energy dissipation in holographic systems is just the energy flux across the bulk black hole horizon: 
	\[
	Q=-\i d^{3}x\left.\sqrt{-g}T_{t}^{z}\right|_{z=1},
	\]
where $T_{N}^{M}$ is the energy momentum tensor. Fig.~\ref{fig:Dissipation-7}
shows the energy dissipation for the whole evolution. We can see that
nearly all nonzero values of $Q$ lies in the period $15<t<250$,
which includes our main process from the initial unstable configurations to the formation of vortex lattices. When $t>250$, the energy dissipation decreases rapidly and goes to $0$ gradually. Fig.~\ref{fig:Dissipation-19} shows energy dissipation when the lattice consists of 19 vortices which shares a similar shape.

\begin{figure}
	\subfloat[\label{fig:Max-Dissipation-Omega}Maximum dissipation as a function of $\Omega$, when $\mu=5.5$. In this plot, the maximum dissipation drops with the increase of $\Omega$.]{\includegraphics[width=0.4\textwidth]{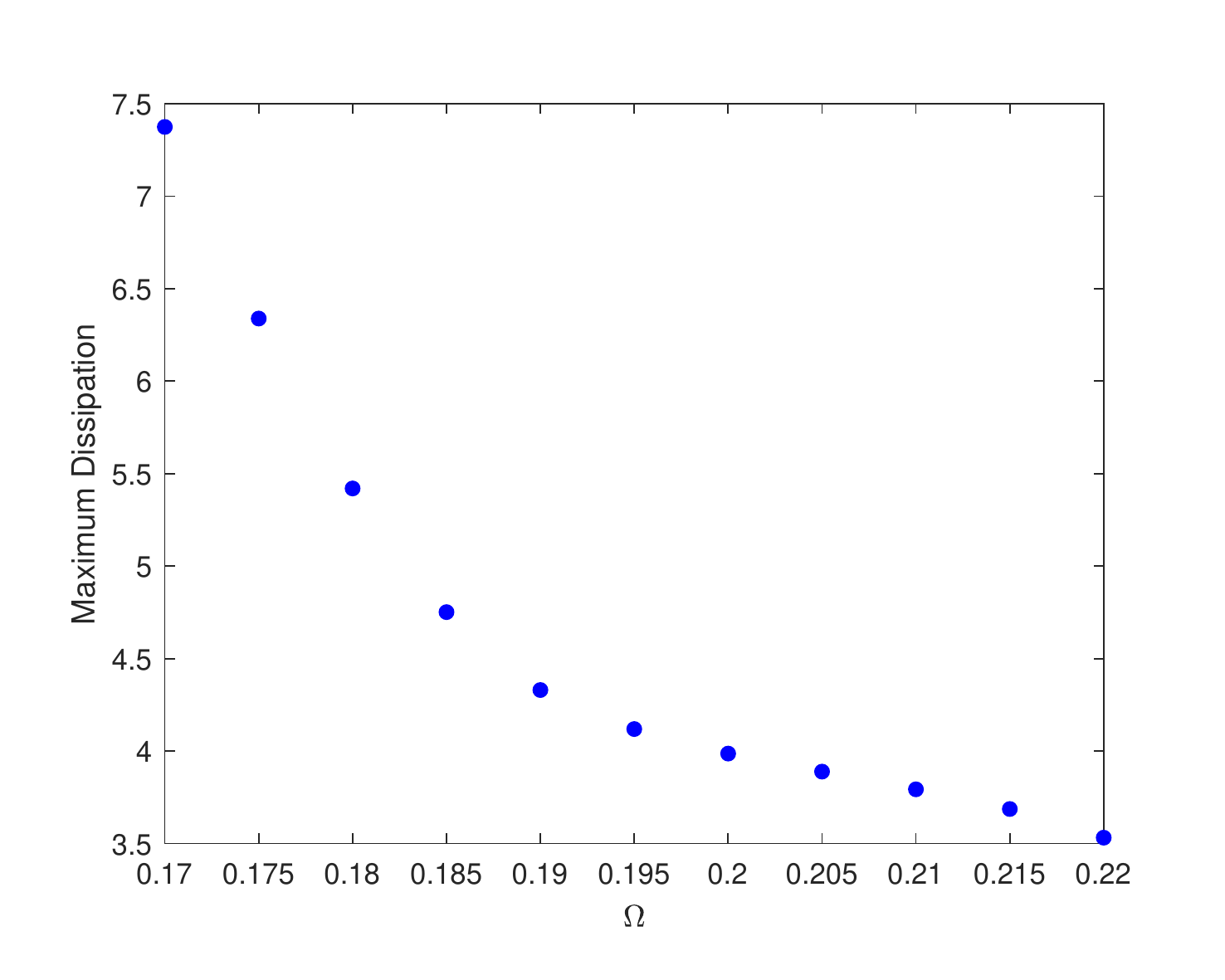}}
	\qquad
	\subfloat[\label{fig:Max-Dissipation-mu}Maximum dissipation as a function of $\mu$, when $\Omega=0.168$. In this plot, the maximum dissipation grows with the increase of $\mu$.]{\includegraphics[width=0.4\textwidth]{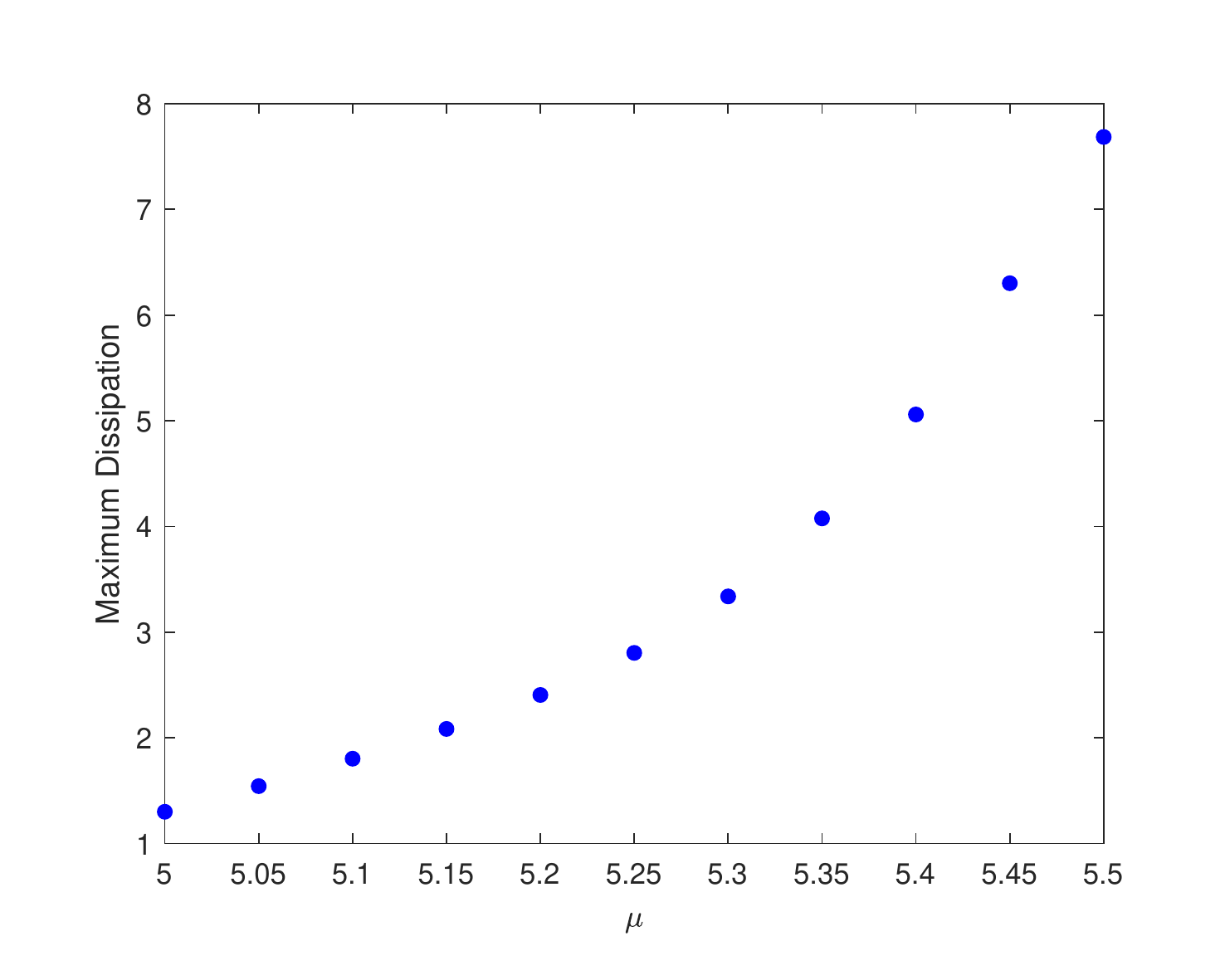}}
	\caption{Maximum dissipation as a function of $\Omega$ and $\mu$, when $R=12$.}\label{fig:Max-Dissipations}
\end{figure}

As shown in Fig.~\ref{fig:Dissipations}, the maximum value of dissipation differs in different evolutions, so it is interesting to study its dependence on other physical quantities, such as $\Omega$ and $\mu$. With other quantities, except for $\Omega$ and $\mu$, fixed, we perform several different evolutions and plot the results in Fig.~\ref{fig:Max-Dissipations}. In Fig.~\ref{fig:Max-Dissipation-Omega}, we fix $\mu=5.5$ and calculate the maximum dissipation when $\Omega$ changes from $0.17$ to $0.22$. In this plot, the maximum dissipation decreases with the increase of $\Omega$. In Fig.~\ref{fig:Max-Dissipation-mu}, we fix $\Omega=0.168$, and calculate the maximum dissipation when $\mu$ increases from $5$ to $5.5$. It is obvious that the maximum dissipation increases with the increase of $\mu$. These results seem to be interesting, but the physical mechanism behind them is yet to be understood further.

\begin{figure}
	\subfloat[\label{fig:Deltat-Omega}$\Delta t$ as a function of $\Omega$, when $\mu=5.5$. In this plot, $\Delta t$ grows roughly with the increase of $\Omega$.]{\includegraphics[width=0.4\textwidth]{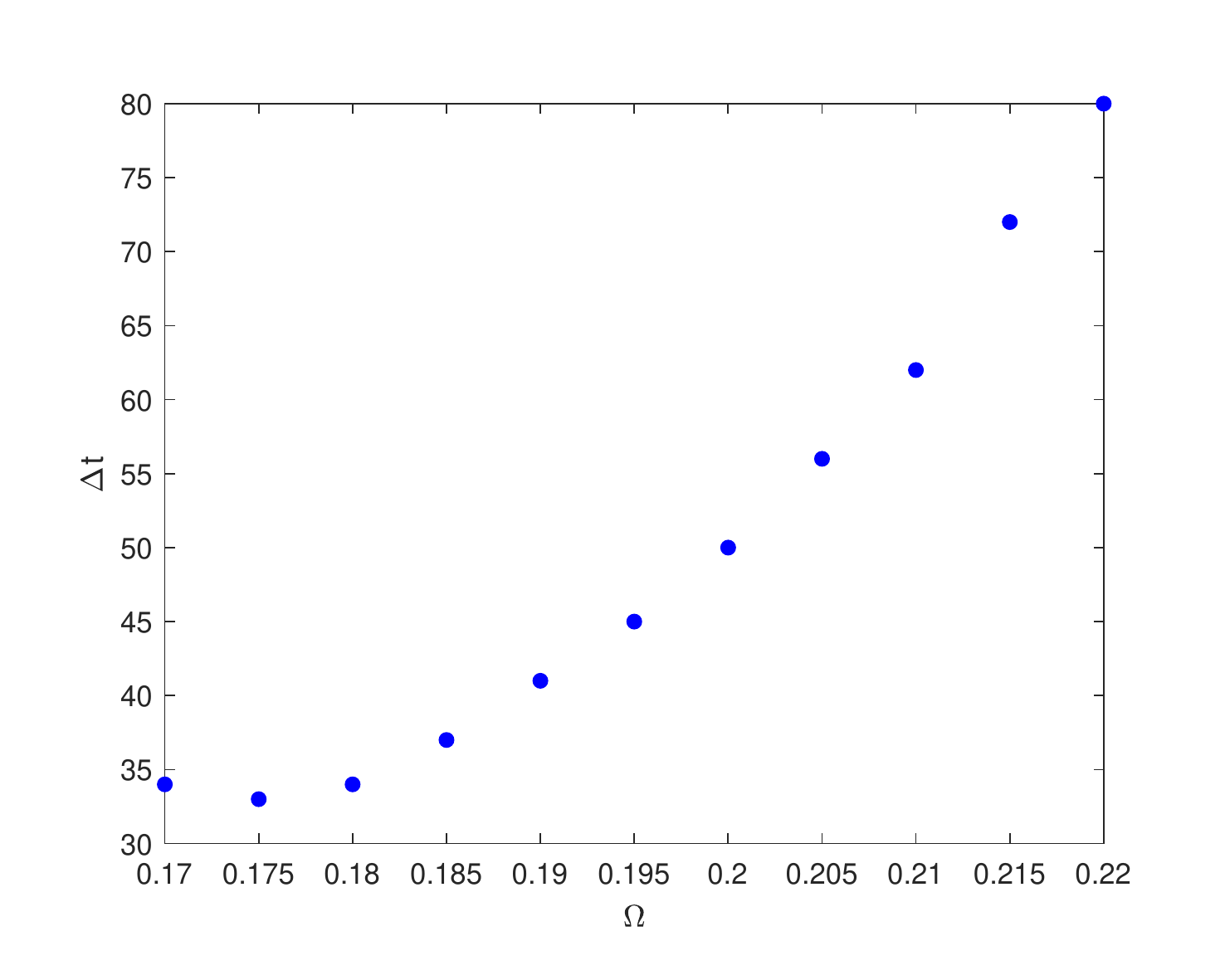}}
	\qquad
	\subfloat[\label{fig:Deltat-mu}$\Delta t$ as a function of $\mu$, when $\Omega=0.168$. In this plot, $\Delta t$ decreses with the increase of $\mu$.]{\includegraphics[width=0.4\textwidth]{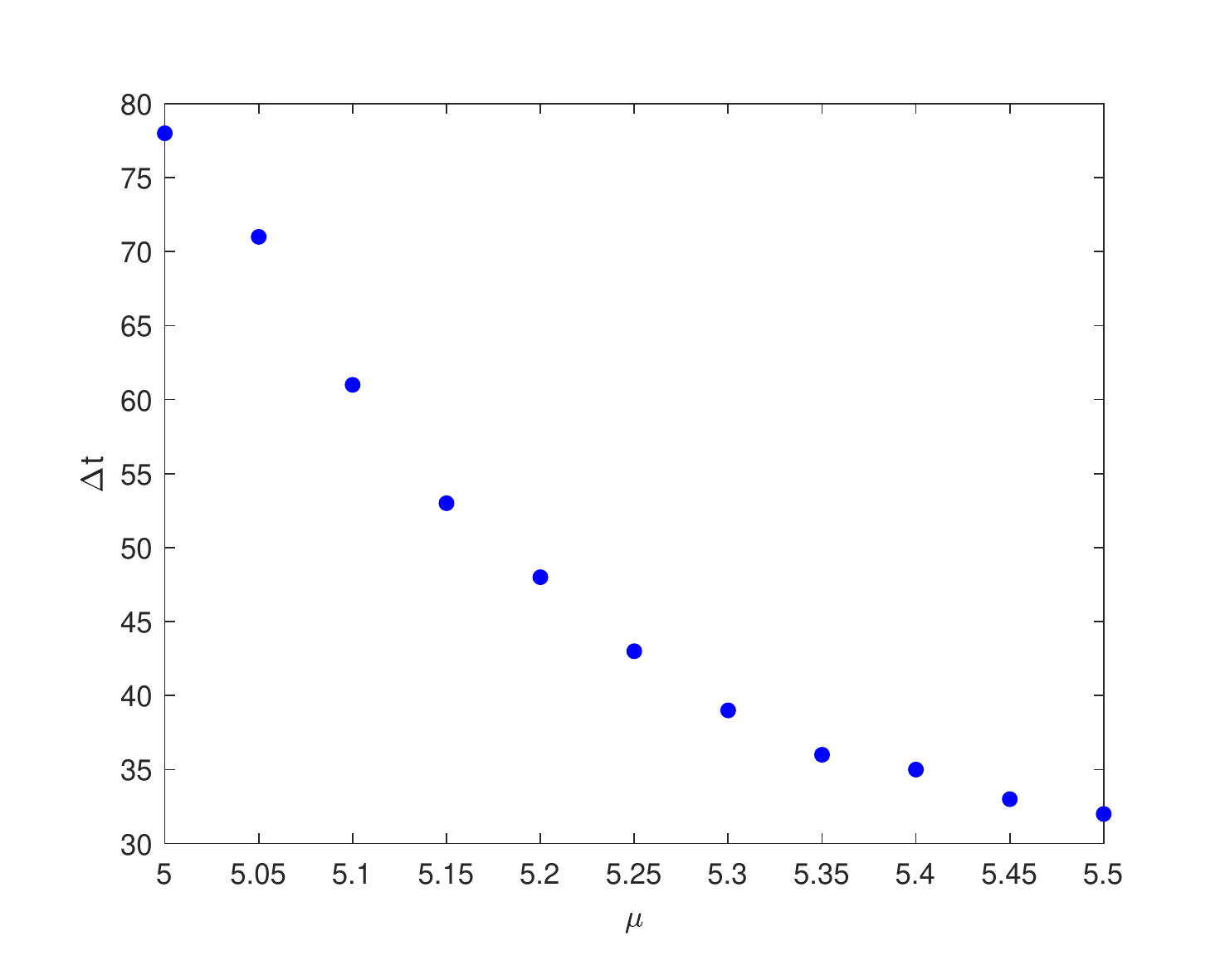}}
	\caption{The period $\Delta t$ between the appearance of the first vortex and the point when dissipation reaches its maximum. And, we choose $R=12$.}\label{fig:Delta_t}
\end{figure}

In our time evolutions, the time when the first vortex appears can be viewed as the start of evolution\footnote{The time before the appearance of the first vortex changes with different amplitudes of perturbations, so is not suitable to describe the system.}, while the time when dissipation reaches its maximum can be regarded to mark the starting point that the system begins to stabilize. As a result, the period between these two moments, which we name as $\Delta t$ from now on, can reflect efficiency of the evolution. Similar to studies of maximum dissipation, we calculate $\Delta t$ when either $\Omega$ or $\mu$ is not fixed, and the results are shown in Fig.~\ref{fig:Delta_t}. In Fig.~\ref{fig:Deltat-Omega}, the tendency that $\Delta t$ grows with the increase of $\Omega$ is witnessed. In Fig.~\ref{fig:Deltat-mu}, $\Delta t$ drops with the increasement of $\mu$, implying that a system with lower temperature involves more quickly. 

\begin{figure}
	\subfloat[\label{fig:Equlibriation_Time-Omega}Equilibriation time $t$ as a function of $\Omega$, when $\mu=5.5$. In this plot, $t$ grows with the increase of $\Omega$ and then remains roughly constant.]{\includegraphics[width=0.4\textwidth]{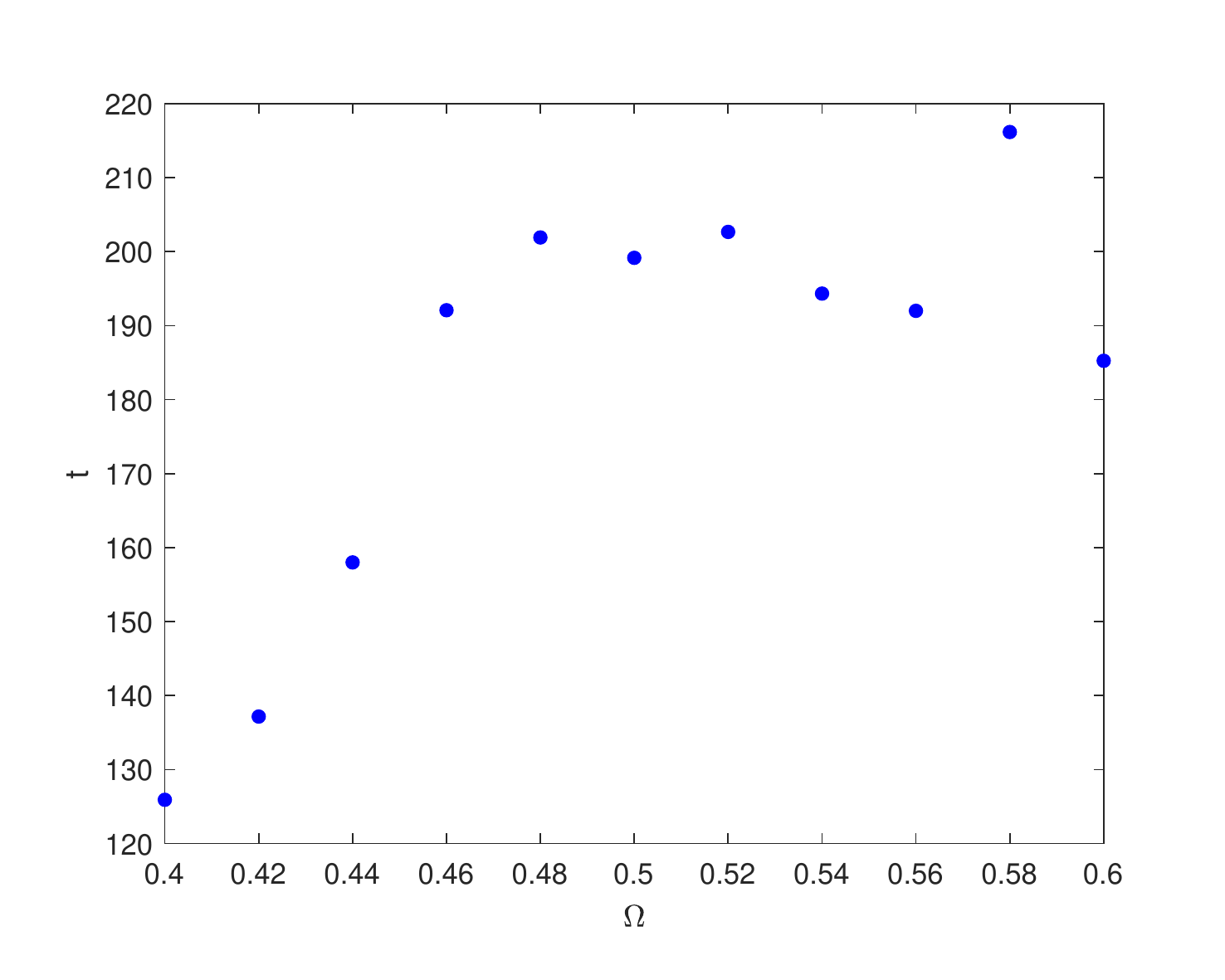}}
	\qquad
	\subfloat[\label{fig:Equlibriation_Time-mu}Equilibriation time $t$ as a function of $\mu$, when $\Omega=0.4$. In this plot, $t$ drops with the increase of $\mu$.]{\includegraphics[width=0.4\textwidth]{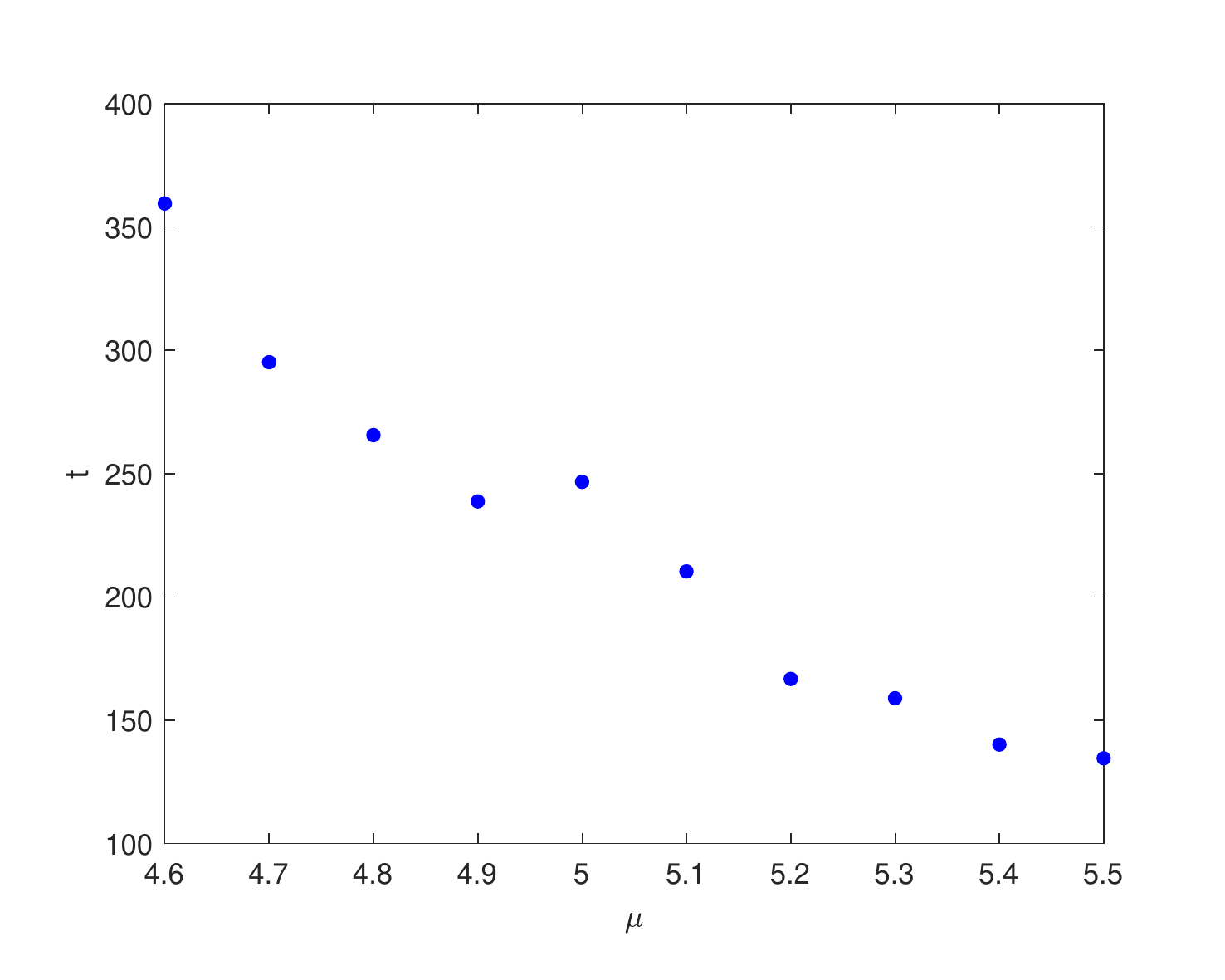}}
	\caption{The equilibriation time when $R=6$.}\label{fig:Equlibriation_Time}
\end{figure}

As discussed in the first paragraph of this section, when lattices stabilize, the dissipations generally reach rather small values, so it is resonable to determine the equilibration time by the period between the moment when the first vortex appears and that when dissipation is small enough. In order to best mimic the true systems, we keep the amplitude of the initial perturbations small, make their phases random, and determine the equlibriation time by the mean equilibriation time of $12$ evolutions\cite{GKLTZ} when $R=6$. Our results are plotted in Fig.~\ref{fig:Equlibriation_Time}. In Fig.~\ref{fig:Equlibriation_Time-Omega}, the equilibriation time grows with the increase of $\Omega$ when $\Omega<0.48$, and then remain roughly constant when $\Omega>0.48$. In Fig.~\ref{fig:Equlibriation_Time-mu}, the equilibriation time drops with the increase of $\mu$, which is similar to the result in Fig.~\ref{fig:Deltat-mu}.

\subsection{Patterns of vortex lattices with 18 vortices}
Vortex lattice has a perfect trangular shape when the number of vortices is infinite, while there are distortions if we consider a finite vortex number case. There are two possibilities. The first case is that the number is so special that interior vortices can form a triangular structure. In this case, there is only one kind of vortex lattice. Fig.~\ref{fig:t=00003D00003D500} for 7 vortices and Fig.~\ref{fig:Lattice-19-500} for 19 vortices are good examples. The second case, in which the number is not so special, is more complicated. Here we will investigate lattices with 18 vortices.

\begin{figure}
	\subfloat[\label{fig:18-1}18-1]{\includegraphics[width=0.2\textwidth]{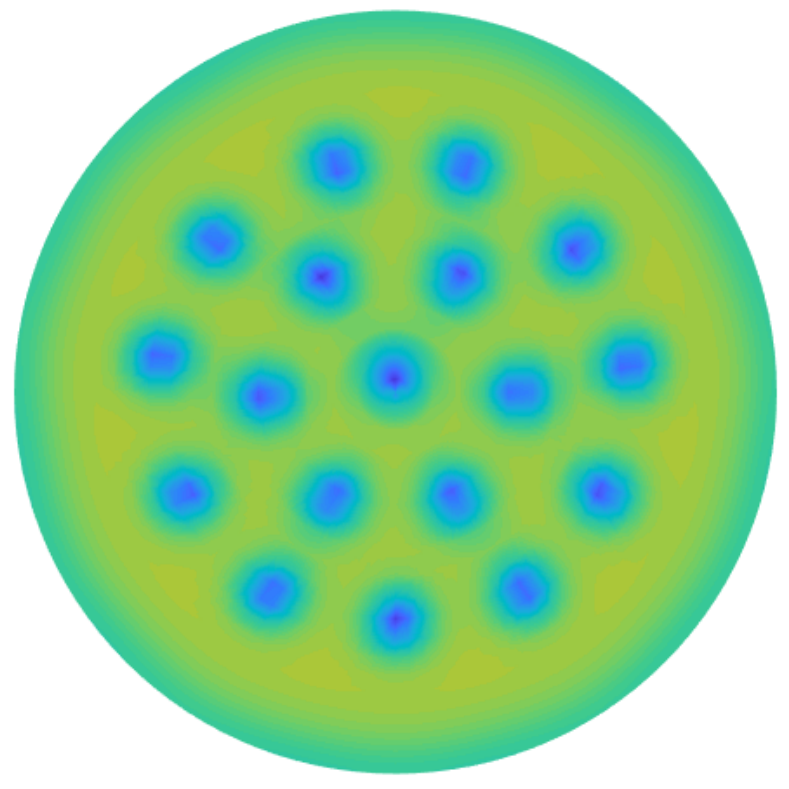}
		
	}\subfloat[\label{fig:18-2}18-2]{\includegraphics[width=0.2\textwidth]{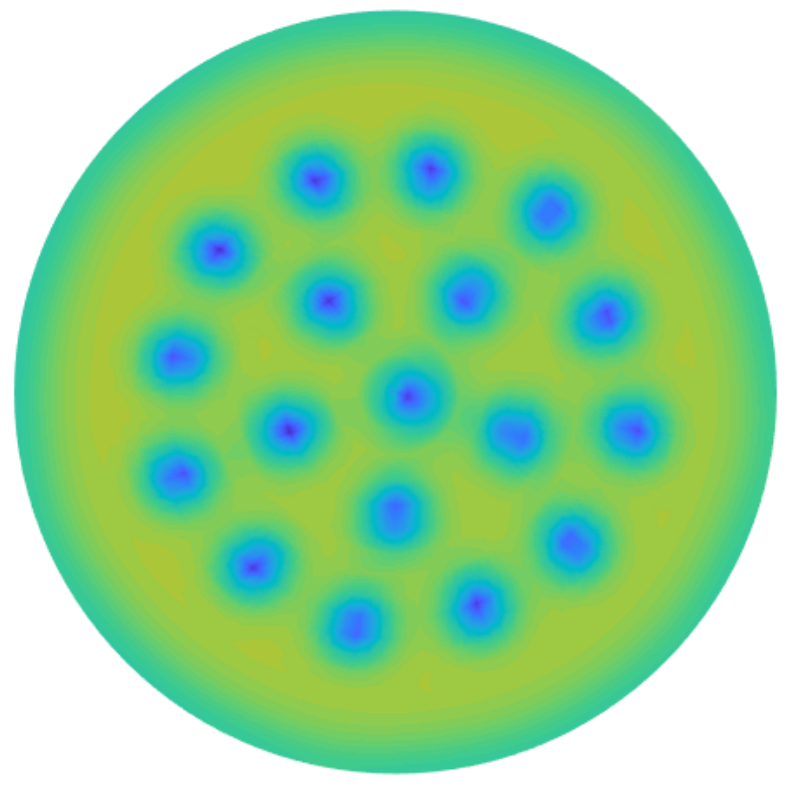}
		
	}\subfloat[\label{fig:18-4}18-4]{\includegraphics[width=0.2\textwidth]{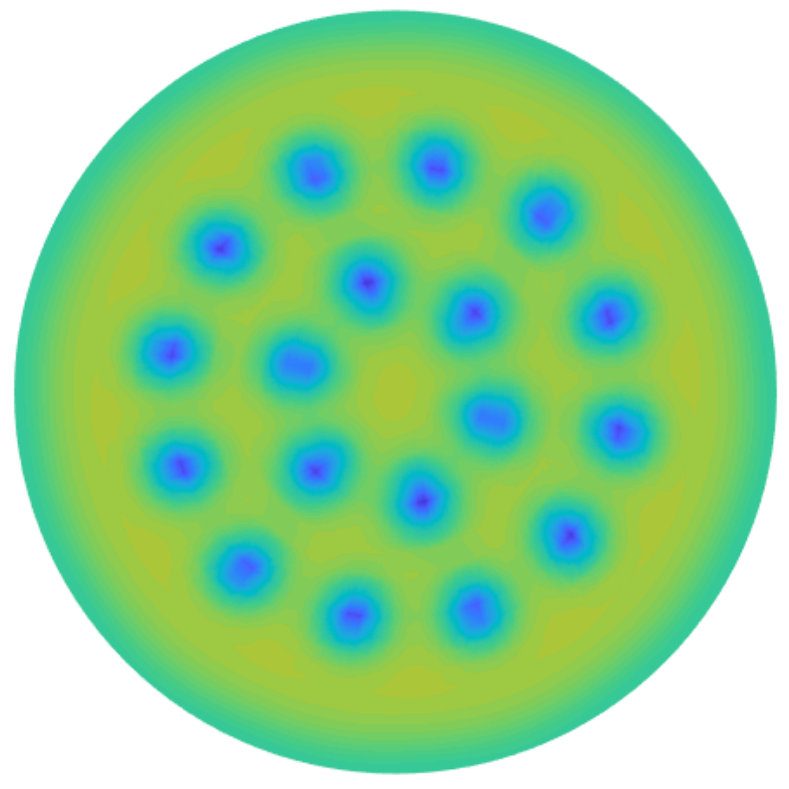}
		
	}\subfloat[\label{fig:18-6}18-6]{\includegraphics[width=0.2\textwidth]{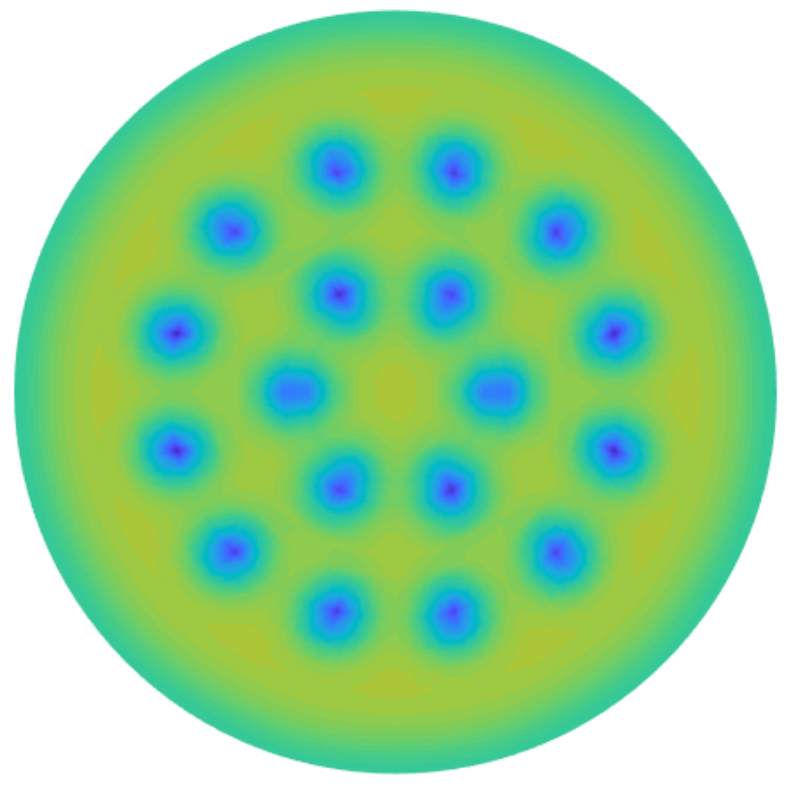}
		
	}\subfloat[\label{fig:18-7}18-7]{\includegraphics[width=0.2\textwidth]{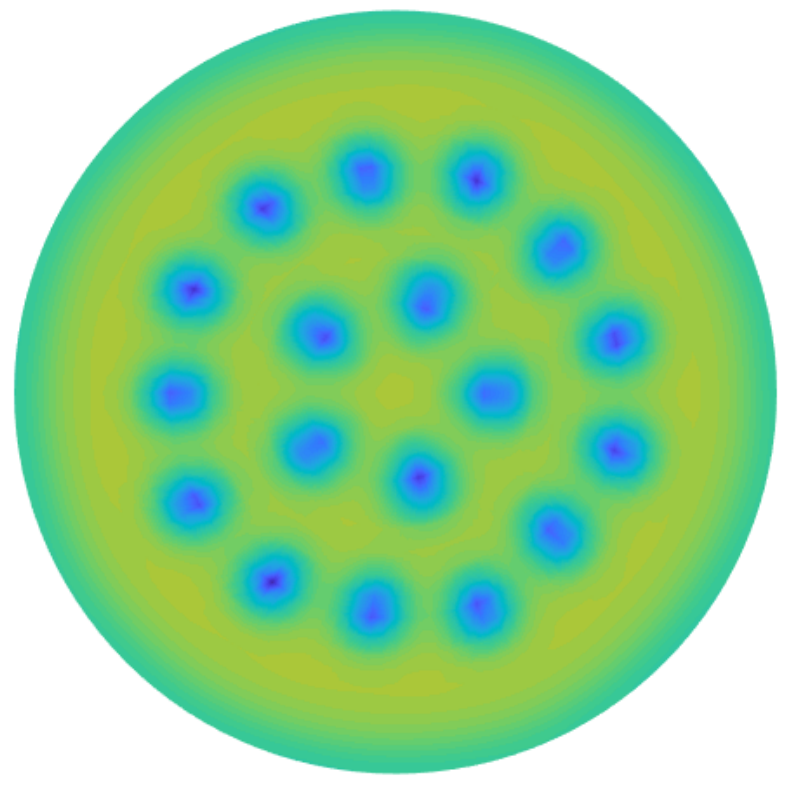}
		
	}\caption{\label{fig:Lattice-18}Condensates as functions of space coordinates for lattices with 18 vortices at $t=500$. In these evolutions, we choose $\protect\Om=0.22$ and $R=12$.}
\end{figure}

It has been calculated that when the lattice consists of 18 vortices, there are 7 different patterns from free energy analyses based on the GP equation \cite{7 Patterns in HeII}. Still, we obtain these lattices from time evolution with perturbations added as equation (\ref{eq:perturbation}). It needs to be emphasized that when $\Om$ is specified, different groups of $A_m$s can result in different evolution processes. In our time evolutions, we obtain 5 patterns with fixed angular velocity (chosen to be $0.22$) but diferent groups of $A_m$s (see Fig.~\ref{fig:Lattice-18}). In this plot, we name these different lattices ``18-N'', where N is the order of the corresponding pattern in \cite{7 Patterns in HeII}.

We have not found the 3rd and 5th patterns from our time evolution.
There exist two possible reasons. First, we differentiate lattices
by comparing shapes of our lattices with results in \cite{7 Patterns in HeII}.
In our model, these 2 absent lattices may not be so distinct that
we can merely recognize them by calculating their free energies, which
cannot be accomplished because of our current numerical accuracy.
The second reason is that these absent lattices are not easy to be
realized from time evolution with random initial perturbations, i.e.
one may need very special initial perturbations to produce them. 

\section{Conclusion and Discussion\label{sec:Conclusion}}

By holography, we have shown that vortices can be generated when the
angular velocity of rotating superfluids exceeds certain critical
values. In numerical simulations of the nonlinear dynamics, these
vortices appear at the edge of the superfluid system first, and then
automatically move into the bulk of the system, where they are eventually
stabilized into certain vortex lattices. For the case of 18 vortices
generated, we have found (at least) five different patterns of the
final lattices formed due to different initial perturbations. Actually,
these patterns can be recognized as five of seven such patterns obtained
from free energy analyses based on the GP equation.

Compared with previous works or traditional methods, our study in
this paper has the following advantages. First, in holography, the
linear instability that can precisely determine the critical angular
velocity is naturally captured by QNM of the bulk black hole. But,
as far as we know, such efficient linear analysis in traditional methods
is still lacking, though it can be done in principle. More importantly,
our study yields all the essential physics of rotating superfluid
systems just using the simplest holographic superfluid model without
any ingredient put by hand, while the traditional methods cannot
incorporate dissipation or finite temperature effects easily and naturally.

For simplicity, we do not consider the backreaction of the matter
fields onto the bulk geometry in our holographic superfluid model.
But for a full holographic duality, such backreaction should be taken
into account\cite{backreaction}. It will be also interesting to
investigate the vortex generation and vortex lattice formation in
holographic superconductors. These topics are left for future exploration. 

\begin{acknowledgments}	
	YT would like to thank Hua-Bi Zeng, Hai-Qing Zhang, Hong Liu and Shan-Quan
	Lan for useful discussions. He would also like to thank the Center
	for Theoretical Physics, Massachusetts Institute of Technology for
	the hospitality. This work is partially supported by NSFC with Grant
	No.11675015 and No.11975235. YT is partially supported by the grants (No. 14DZ2260700)
	from Shanghai Key Laboratory of High Temperature Superconductors. He is also supported by the ``Strategic
	Priority Research Program of the Chinese Academy of Sciences'' with
	Grant No.XDB23030000. HZ is also supported by the Vrije Universiteit
	Brussel through the Strategic Research Program ``High-Energy Physics'',
	and he is also an individual FWO Fellow supported by 12G3515N. 
\end{acknowledgments}

\begin{appendix}
	
\section{Equations for Quasi-Normal Modes}\label{sec:EOMs_for_QNMs}
We solve problems of quasi-normal modes in Eddington-Finkelstein coordinates, so metric (\ref{eq:metric}) is adopted. Linealize equations (\ref{eq:evolve_psi}-\ref{eq:evolve_Ah}), substitute  (\ref{eq:perturbation_of_psi}) and (\ref{eq:perturbation_of_A}) into the linealized equations, then we reach the final form of equations for quasi-normal modes:
\begin{align}
	\left[-2i\left(\omega+A_{t}\right)\p_{z}-i\p_{z}A_{t}-\p_{z}\left(f\p_{z}\right)-\left(\p_{r}-iA_{r}\right)^{2}-\frac{\p_{r}-iA_{r}}{r}+\frac{\left(A_{\t}-m\right)^{2}}{r^{2}}+z\right]p\nonumber \\
	-\left(i\s\p_{z}+2i\p_{z}\s\right)a_{t}+\left(i\s\p_{r}+2i\p_{r}\s+2\s A_{r}+\frac{i\s}{r}\right)a_{r}+\frac{\s}{r^{2}}\left(2A_{\t}-m\right)a_{\t} & =0,\\
	\left[-2i\left(\omega-A_{t}\right)\p_{z}+i\p_{z}A_{t}-\p_{z}\left(f\p_{z}\right)-\left(\p_{r}+iA_{r}\right)^{2}-\frac{\p_{r}+iA_{r}}{r}+\frac{\left(A_{\t}-m\right)^{2}}{r^{2}}+z\right]q\nonumber \\
	+\left(i\s^{*}\p_{z}+2i\p_{z}\s^{*}\right)a_{t}-\left(i\s^{*}\p_{r}+2i\p_{r}\s^{*}-2\s^{*}A_{r}+\frac{i\s^{*}}{r}\right)a_{r}+\frac{\s^{*}}{r^{2}}\left(2A_{\t}+m\right)a_{\t} & =0,\\
	\left(-i\s^{*}\p_{z}+i\p_{z}\s^{*}\right)p+\left(i\s\p_{z}-i\p_{z}\s\right)q+\p_{z}^{2}a_{t}-\left(\frac{\p_{z}}{r}+\p_{z}\p_{r}\right)a_{r}-\frac{im}{r^{2}}\p_{z}a_{\t} & =0,\\
	\left[if\left(\s^{*}\p_{z}-\p_{z}\s^{*}\right)-\left(\o+2A_{t}\right)\s^{*}\right]p+\left[if\left(\p_{z}\s-\s\p_{z}\right)+\left(\omega-2A_{t}\right)\s\right]q\nonumber \\
	+\left(i\omega\p_{z}+\p_{r}^{2}+\frac{1}{r}\p_{r}-2\left|\s\right|^{2}-\frac{m^{2}}{r^{2}}\right)a_{t}+\left(f\p_{z}+i\o\right)\left(\p_{r}+\frac{1}{r}\right)a_{r}+\frac{m}{r^{2}}\left(if\p_{z}-\o\right)a_{\t} & =0,\\
	\left(i\s^{*}\p_{r}-i\p_{r}\s^{*}+2\s^{*}A_{r}\right)p+\left(-i\s\p_{r}+i\p_{r}\s+2\s A_{r}\right)q\nonumber \\
	-\p_{z}\p_{r}a_{t}+\left(-2i\omega\p_{z}-f\p_{z}^{2}-f'\p_{z}+\frac{m^{2}}{r^{2}}+2\left|\s\right|^{2}\right)a_{r}+\frac{im}{r^{2}}\p_{r}a_{\t} & =0,\\
	\left(2A_{\t}-m\right)\s^{*}p+\left(2A_{\t}+m\right)\s q-im\p_{z}a_{t}+im\left(\p_{r}-\frac{1}{r}\right)a_{r}\nonumber \\
	+\left[-2i\omega\p_{z}-\p_{z}\left(f\p_{z}\right)-\p_{r}^{2}+\frac{\p_{r}}{r}+2\left|\s\right|^{2}\right]a_{\t} & =0.
\end{align}
Note that the above equations are a set of coupled equations for $(p,q,a_{\mu})$,
coming from the $e^{-i\o t+im\t}$ terms.
	
\section{Equations of Motion for Time Evolution}\label{sec:EOMs_for_Evolution}
In order to perform time evolution, Eddington-Finkelstein
coordinates are adopted. Substitute metric (\ref{eq:metric}) into equations (\ref{eq:K-G}) and (\ref{eq:Maxwell}), the following equations of motion are obtained
\begin{align}
	-\left[\p_{z}\left(f\p_{z}\s\right)+i\left(\p_{z}A_{t}\right)\s+2iA_{t}\p_{z}\s\right]+\left[-\p_{r}^{2}\s+i\left(\p_{r}A_{r}\right)\s+2iA_{r}\p_{r}\s-\frac{1}{r}\p_{r}\s\right]\nonumber \\
	+\frac{1}{r^{2}}\left[-\p_{\t}^{2}\s+i\left(\p_{\t}A_{\t}\right)\s+2iA_{\t}\p_{\t}\s\right]+\left[A_{r}^{2}+\frac{iA_{r}}{r}+\frac{A_{\t}^{2}}{r^{2}}+z\right]\s+2\p_{t}\p_{z}\s & =0,\label{eq:evolve_psi}\\
	\p_{z}^{2}A_{t}-\p_{z}\p_{r}A_{r}-\frac{1}{r}\p_{z}A_{r}-\frac{1}{r^{2}}\p_{z}\p_{\t}A_{\t}-i\left(\s^{*}\p_{z}\s-\s\p_{z}\s^{*}\right) & =0,\label{eq:restrict}\\
	-\frac{1}{r}\p_{t}A_{r}-\p_{t}\p_{z}A_{t}-\p_{t}\p_{r}A_{r}-\frac{1}{r^{2}}\p_{t}\p_{\t}A_{\t}+\frac{f}{r}\p_{z}A_{r}+f\p_{z}\p_{r}A_{r}+\frac{f}{r^{2}}\p_{z}\p_{\t}A_{\t}+\p_{r}^{2}A_{t}\nonumber \\
	+\frac{1}{r}\p_{r}A_{t}+\frac{1}{r^{2}}\p_{\t}^{2}A_{t}-i\left(\s^{*}\p_{t}\s-\s\p_{t}\s^{*}\right)-2A_{t}\s^{*}\s+if\left(\s^{*}\p_{z}\s-\s\p_{z}\s^{*}\right) & =0,\label{eq:boudary_rho}\\
	2\p_{t}\p_{z}A_{r}-\p_{z}\p_{r}A_{t}-\p_{z}\left(f\p_{z}A_{r}\right)+\frac{1}{r^{2}}\left(\p_{r}\p_{\t}A_{\t}-\p_{\t}^{2}A_{r}\right)\nonumber \\
	+i\left(\s^{*}\p_{r}\s-\s\p_{r}\s^{*}\right)+2A_{r}\s^{*}\s & =0,\label{eq:evolve_Ar}\\
	2\p_{t}\p_{z}A_{\t}-\p_{z}\left(f\p_{z}A_{\t}\right)-\p_{z}\p_{\t}A_{t}-\p_{r}^{2}A_{\t}+\frac{1}{r}\p_{r}A_{\t}+\p_{r}\p_{\t}A_{r}-\frac{1}{r}\p_{\t}A_{r}\nonumber \\
	+i\left(\s^{*}\p_{\t}\s-\s\p_{\t}\s^{*}\right)+2A_{\t}\s^{*}\s & =0.\label{eq:evolve_Ah}
\end{align}

\end{appendix}

\end{document}